\documentclass[twocolumn]{aastex62}

\usepackage{amsmath}
\usepackage[inline]{enumitem}
\usepackage{natbib}
\usepackage{array,multirow}

\newcommand{\dgs}{$^\circ$ }

\received{2018 November 12}
\revised{2019 February 6}
\accepted{2019 March 14}
\submitjournal{AJ}

\shorttitle{Observing Rotation Rates on Eccentric Aquaplanets}
\shortauthors{Adams, Boos \& Wolf}

\begin{document}

\title{Aquaplanet Models on Eccentric Orbits --- Effects of Rotation Rate on Observables}

\author[0000-0002-7139-3695]{Arthur D. Adams}
\affiliation{Department of Astronomy, Yale University, New Haven, CT 06520}
\author[0000-0001-9076-3551]{William R. Boos}
\affiliation{Department of Earth and Planetary Science, University of California, Berkeley, CA 94720}
\affiliation{Climate and Ecosystem Sciences Division, Lawrence Berkeley National Laboratory, Berkeley, CA 94720}
\author[0000-0002-7188-1648]{Eric T. Wolf}
\affiliation{Laboratory for Atmospheric and Space Physics, Department of Atmospheric and Oceanic Sciences, University of Colorado, Boulder, CO 80309}

\correspondingauthor{Arthur D. Adams}
\email{arthur.adams@yale.edu}

\begin{abstract}
Rotation and orbital eccentricity both strongly influence planetary climate. Eccentricities can often be measured for exoplanets, but rotation rates are currently difficult or impossible to constrain. Here we examine how the combined effects of rotation and eccentricity on observed emission from ocean-rich terrestrial planets can be used to infer their rotation rates in circumstances where their eccentricities are known. We employ an Earth climate model with no land and a slab ocean, and consider two eccentricities ($e=0.3$ and 0.6) and two rotation rates: a fast Earth-like period of 24 hours, and a slower pseudo-synchronous period that generalizes spin synchronization for eccentric orbits. We adopt bandpasses of the Mid-Infrared Instrument on the James Webb Space Telescope as a template for future photometry. At $e=0.3$ the rotation rates can be distinguished if the planet transits near periastron, because slow rotation produces a strong day-night contrast and thus an emission minimum during periastron. However, light curves behave similarly if the planet is eclipsed near periastron, as well as for either viewing geometry at $e=0.6$.  Rotation rates can nevertheless be distinguished using ratios of emission in different bands, one in the water vapor window with another in a region of strong water absorption. These ratios vary over an orbit by $\lesssim\!0.1$ dex for Earth-like rotation, but by 0.3--0.5 dex for pseudo-synchronous rotation because of large day-night contrast in upper-tropospheric water. For planets with condensible atmospheric constituents in eccentric orbits, rotation regimes might thus be distinguished with infrared observations for a range of viewing geometries.
\end{abstract}


\section{Introduction}\label{sec:introduction}
The geometry of a planet's orbit and the rate of its rotation are both key to understanding spatial and temporal variations in the heating of its atmosphere by its host star. On Earth, the solar heating at a given position and time is dominated by the 24-hour day-night cycle from rotation, as well as the annual cycle in the orientation of the axial tilt with respect to the Sun-Earth line. These diurnal and annual cycles, with their significantly different time scales, operate largely independently of each other. Additionally, Earth's nearly circular orbit means that variations in the Earth-Sun distance are small; Earth's seasons are driven primarily by obliquity rather than eccentricity.  Differences in orbital eccentricity, planetary rotation rate, and axial tilt can all have large consequences for atmospheric heating rates and planetary climate. Here we examine how rotation rate and orbital eccentricity control, via the global atmospheric circulation, the radiative properties of a planet's surface and atmosphere. By using detailed representations of the circulation and radiative transfer, we wish to explore whether the rough scale of rotation rate can be inferred from a limited set of observable features.

In order to extend our understanding of how planetary rotation and orbit drive periodicities in temperatures on other planets, we need to obtain observational constraints on actual rotation rates and orbits. In cases where exoplanets have been detected via both the radial velocity and transit methods, one can reliably constrain both the orbital periods and eccentricities to fairly high precision. However, very few observational constraints exist for the rotation rates of exoplanets, Earth-like or otherwise. From tidal arguments we expect Hot Jupiters on short orbital periods to undergo spin-orbit synchronization\footnote{This term is often used interchangeably with tidal locking. They are equivalent for circular orbits, but a perpetual day side scenario is not strictly possible for nonzero orbital eccentricities due to the non-constant rate of change in true anomaly over the orbit.} on timescales shorter than the ages of the systems \citep{gol66,sho02}. More recently, \citet{dew16} and \citet{lew17} used phase photometry to constrain the range of possible rotation periods for the highly eccentric giant planet HD 80606 b. No observational constraints yet exist for the rotation rate of Earth-sized exoplanets.

The rotation rate will set both the motion of the sub-stellar point on the planet's surface and the nature of the global atmospheric circulation that, in turn, controls planetary climate. \citet{mer10} demonstrated that for Earth-like planets on circular orbits, Earth-like rotation periods will have larger latitudinal gradients in temperature away from the equator \citep[see also][]{Cullum2014}. In contrast, when rotation is slow ($P_{\mathrm{rot}}\!\sim\!P_{\mathrm{orb}}$) the surface temperature should scale with the local instellation, peaking at or near the longitude of the sub-stellar point, with weaker equator-to-pole temperature differences. At the surface, the effect of rotation rate on temperature should also depend on the depth of the ocean mixed layer. \citet{bol16} explored a range of eccentricities for aquaplanet models and demonstrated that increasing the thermal inertia of the oceans damps changes in the climate more efficiently. Accordingly, faster rotation in their models reduces the sensitivity of the climate to the thermal inertia of the oceans.

When planets on highly non-circular orbits are considered, we expect many of the general predictions based on the broad regimes of rotation rate for circular orbits to be extensible. In the limit of slow rotation, the pseudo-synchronous rotation rate \citep{hut81} effectively approximates spin-orbit synchronization around the time of periastron, when stellar forcing is maximal. The physicality of this predicted rate relies on certain assumptions, including a constant tidal lag, and its applicability to terrestrial planets is still debated. \citet{mak13} make the case that the only stable equilibrium states for terrestrial systems are spin-orbit resonances, for example the 3:2 spin-orbit resonance seen in Mercury. Acknowledging this, in the calculations performed herein we adopt pseudo-synchronization primarily as an example of a rotation rate which for most eccentricities will be much slower than an Earth-like rotation. In contrast, for a much faster rotation like Earth's, we expect that strong east-west winds induced by planetary rotation will homogenize temperatures in longitude on time scales shorter than the rate of change in instellation due to the eccentric orbit\footnote{For a review of the circulation of Earth's atmosphere, see \citet{sch06}. For a recent review of the circulation of tidally-locked planets, see \citet{Pierrehumbert2019}.}.

With current observational limitations in mind, here we seek to understand the time variation of surface and atmospheric temperatures on Earth-like planets with contrasting rotation rates and orbital shapes. Our goal is to determine whether the effects of rotation and orbit on incident radiation could induce a response which would lead to observable differences, thereby indirectly providing a method for estimating the rotation rate when only eccentricity is constrained a priori. In the scenarios explored here we assume zero obliquity; unlike Earth, the primary driver of seasonal variations will be eccentricity rather than axial tilt. To do so we adapt a class of 3-D models, often referred to as general circulation models (GCMs), which are most often used to simulate Earth's atmosphere. These GCMs allow for analysis of the effects of various properties of the planetary system on the evolution of climate \citep[e.g.][]{ogo08,wol13,wol14,wol15}. Such models have also been developed for other terrestrial planets in the Solar System, particularly Venus and Mars \citep[e.g.][]{ros83,bar93,bar96a,hab93,bar96b,for99,leb10,zal10,for13}, and have been used to explore large-scale atmospheric circulation under differences in atmospheric composition, rotation rate, and surface gravity.

Beyond the Solar System, GCMs now have a considerable history of use for possible planetary scenarios in other stellar systems. Significant modifications have been undertaken by numerous groups to accommodate unfamiliar orbital and surface conditions. \citet{jos97} present an early example of a GCM applied to a hypothetical extrasolar planet, exploring the consequences of putting an Earth-like planet on a short-period, spin-synchronous orbit around a late-type star. Following this, \citet{mer10} used a GCM to model an ``aquaplanet'', an Earth-like planet with its entire surface covered by water, with results discussed above. Such works adopt the complex representations of physics operating below the GCM grid scale (e.g.\ precipitating atmospheric convection and radiative transfer), originally developed for Earth, to predict the behavior of exoplanet atmospheres.

In both \citet{jos97} and \citet{mer10}, the authors assume a stable ocean cover for their initial conditions. The validity of this assumption is studied explicitly through the definition and continual refinement of the Habitable Zone (HZ), which is the range of star-planet separations for which a planet with an Earth-like mass, radius, and atmospheric composition and surface pressure can plausibly support liquid water oceans on its surface. The foundational works for our current HZ definition rely primarily on a 1-D radiative-convective climate model \citep{kas93,kop13}. These idealized 1-D models make some assumptions most relevant for Earth, but subsequent efforts by a wide range of researchers have extended the range of theoretical habitability, both in model complexity and parameter space \citep[see review by][]{Ramirez2018}. \citet{Leconte2013} demonstrated that 3-D modeling is necessary to account for large-scale surface temperature contrasts due to inefficient or non-isotropic energy redistribution; these conditions expand the range of stellar fluxes where a runaway greenhouse effect may be prevented. Also using 3-D modeling, \citet{yan13} showed that accounting for cloud distribution and dynamics further expanded the parameter space where climates could be habitable, particularly for slowly rotating planets. More recently, \citet{Way2018}  explored the effects of ocean heat transport, which is known to be a significant mechanism for energy transport on Earth, for a range of rotation rates and stellar fluxes.

Most relevant to our study are works that explored effects of orbital eccentricity, such as the interplay of obliquity and eccentricity using climate models of intermediate complexity \citep{lin15}, and using GCMs to explore the possible limitations of assumptions for sustained ocean cover \citep{bol16}, as well as non-terrestrial atmospheric properties and stellar types \citep{shi16}. GCMs have also been used under similar water-rich conditions to study changes in stellar luminosity due to spectral types \citep{shi13,shi14} or evolution \citep{wol15}, and the work of \citet{Ramirez2016} complements this analysis with a combination of stellar and planetary mass loss models, accounting for physical mechanisms that may be relevant for potentially habitable planets orbiting a variety of host stellar types. Taken together, these previous studies provide a wide range of theoretical predictions for the conditions governing the possible existence and persistence of liquid water oceans on terrestrial-size planets. While our purpose is not to make a critique or refinement of the currently defined HZ, and while we limit ourselves to a constant luminosity, solar-type host star, we introduce these studies here to place a precedent for the range of orbital configurations currently thought to support ocean-covered planets.

The two primary parameters we vary are orbital eccentricity and rotation period, both of which have undergone substantial study in recent literature, especially for close-in giant planets. A significant amount of work has been done to model the atmospheric response of highly eccentric Hot Jupiters, given their much more favorable observability when compared with planets on Earth-like orbits. \citet{Langton2008} made foundational hydrodynamic simulations of the upper atmospheres of known Hot Jupiters with orbital eccentricities as high as $e=0.93$ (HD 80606 b), and demonstrated that the intense stellar forcing during periastron passage was the primary driver of atmospheric dynamics. For a similar set of eccentric planets, \citet{cow11a} characterized the predicted phase variations from the orbital and viewing geometry and bulk atmospheric dynamics and radiative transfer, an analysis which was extended to predict thermal timescales for planets on Earth-like orbits in \citet{Cowan2012}. \citet{Kataria2013} presented the first fully 3-D simulations for eccentric Hot Jupiters, and incorporated both the mean-flux normalization and pseudo-synchronous rotation assumptions that we adopt here. They found that planets on both eccentric and circular orbits exhibit qualitatively similar atmospheric features, such as equatorial jets and day-night temperature differences. These features depend largely on the rotation rate, which sets the strength of the Coriolis forces. Furthermore, they demonstrated that the viewing geometry, in particular the longitude of periastron, has a major effect on the observed shape and offset of thermal phase variations.

While GCMs can be used to explore a range of interesting and hypothetical atmospheric dynamics that might occur on exoplanets, we would like to go beyond this to make testable predictions. Here we focus on predictions that might be verified through broadband photometry, which generally offers a greater photon count over spectroscopy and therefore is invaluable for studying small, warm-to-cool, and therefore faint exoplanets. Many giant planets on extremely short orbits have been observed indirectly via transits and secondary eclipses, and in some cases have been examined over significant fractions of their orbits [see recent reviews by \citet{par17} and \citet{Kreidberg2018}]. While transit detections have substantially increased the population of known extrasolar planets, secondary eclipse measurements provide a complementary set of data that helps constrain major properties of planets' emissions. From the depth of a planet's eclipse we can infer the temperature of its illuminated hemisphere, which gives clues to the atmospheric conditions. A key instrument for observing secondary eclipses has been the Spitzer Space Telescope's Infrared Array Camera  (IRAC) \citep{wer04}, which has 4 photometric bands spanning 3.6--8.0 $\mu$m. The majority of Spitzer secondary eclipses and phase curves observed with Spitzer were made during the ``warm Spitzer'' phase, with only the 3.6 and 4.5 $\mu$m remaining operational; a summary of these measurements and references can be found in \citet{ada18b}. In some opportune cases Spitzer has been able to observe planets over full orbits in some combination of these bands, providing a temporal connection between night-side observations of a planet in transit and day-side observations of it in eclipse.

Taking inspiration from these past analyses, as well as current observational techniques, here we generate predictions of eclipse depths and phase photometry from an exoplanet GCM. We begin by describing our assumptions for both orbital eccentricity and planetary rotation rate, along with relevant background, in \S \ref{sec:theory}. In \S \ref{sec:model} we describe the GCM we employ, and how we use it to simulate observable quantities. We present our results in \S \ref{sec:results}, focusing on both the internal properties of the planetary atmospheres and the consequent observables.

\section{Assumptions of Rotation and Orbit}\label{sec:theory}

\subsection{Orbital Eccentricity}\label{sec:theory:eccentricity}
Numerous HZ planets are known to have nonzero orbital eccentricity \citep{ada16}. The role of the Lidov-Kozai mechanism \citep{koz62,lid62} for increasing the orbital eccentricity of terrestrial planets is explored in \citet{spi10}, and subsequently in works such as \citet{Georgakarakos2016}, \citet{Way2016}, and \citet{Deitrick2018}. The process allows for the existence of highly eccentric Earth-sized planets with neighboring giant planets; orbital resonances between the planets and their mutual proximities are the two critical components to the mechanism's efficiency \citep{1999ssd..book.....M}.

\citet{wil02} argue that the instellation time-averaged over an orbit is the primary determinant of whether liquid water can be sustained on a terrestrial planet's surface. For eccentric orbits the mean-flux approximation (MFA) fixes the time-averaged instellation over an orbit to that of a reference planet on a circular orbit. When comparing the flux $F$ with that of Earth, as a function of the stellar luminosity $L_\star$, semi-major axis $a$, and eccentricity $e$, 
\begin{equation}\label{eq:MFA}
\left\langle \frac{F}{F_\oplus} \right\rangle = \frac{L_\star/L_\odot}{\left(a/a_\oplus\right)^2 \sqrt{1 - e^2}}
\end{equation}
where the reference values for Earth are taken to be $F_\oplus=1360$ W m$^{-2}$, $L_\odot=3.83\times10^{26}$ W, and $a_\oplus=1$ AU.  Triangle brackets denote a time average over an orbit.  \citet{bar08} use the conclusion of \citet{wil02} to define an ``Eccentric HZ'' (EHZ) by scaling the semi-major axis $a$ with eccentricity according to the MFA (Equation \ref{eq:MFA}). All models presented here have their semi-major orbital axes (and, by extension, orbital periods) set according to this approximation.

\subsection{Rotation Rate}\label{sec:theory:rotation}

\begin{figure}[htb!]
\begin{center}
    \includegraphics[width=8.5cm]{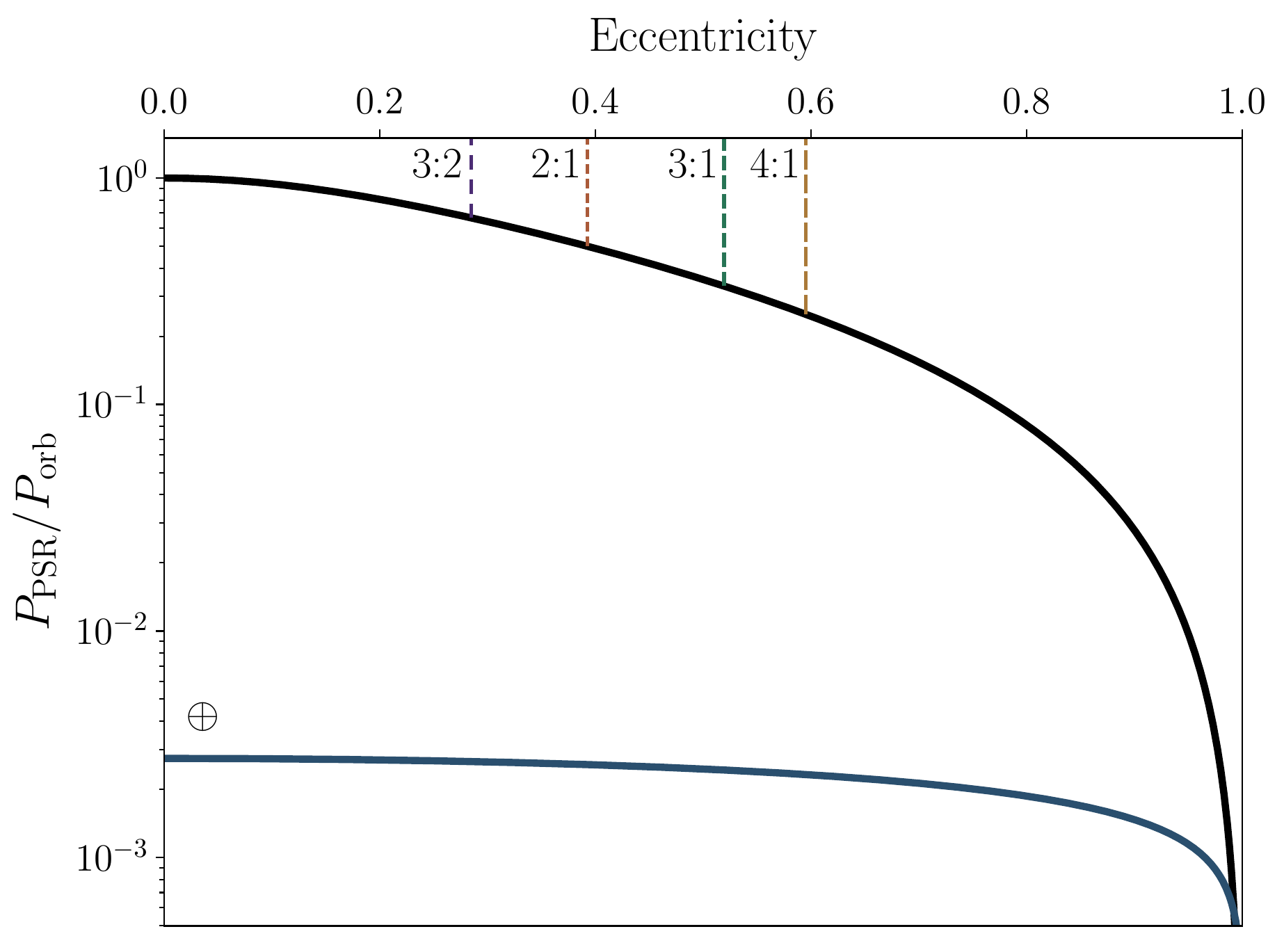}
\caption{The theoretical pseudo-synchronous rotation period matches the synchronous period (orbital period) for a circular orbit, and remains on the order of the orbital period until very high eccentricity, where the ratio (shown in black) drops precipitously as $e \rightarrow 1$. Here we assume the orbital period scales with the Mean-Flux Approximation (Equation \ref{eq:MFA}), which preserves the orbit-integrated instellation as eccentricity is changed. The ratio corresponding to a rotation period of 1 Earth day is shown as the curve in blue, labelled as $\oplus$. The values of eccentricity corresponding to spin-orbit resonances of 3:2 ($e=0.285$), 2:1 ($e=0.392$), 3:1 ($e=0.519$), and 4:1 ($e=0.595$) are marked.}
\label{fig:PSR_ratio}
\end{center}
\end{figure}

Analogous to the synchronous limit for circular orbits, \citet{hut81} presents a limiting rotation rate based on a tidal evolution argument for binary systems with eccentric orbits. The pseudo-synchronous rotation (PSR) period is calculated from this pseudo-synchronous rate, and may be written in units of the orbital period as
\begin{equation}\label{eq:PSR}
\frac{P_{\mathrm{PSR}}}{P_{\mathrm{orb}}} = \frac{\left( 1 + 3 e^2 + \frac{3}{8} e^4 \right) \left(1 - e^2 \right)^{3/2}}{1 + \frac{15}{2} e^2 + \frac{45}{8} e^4 + \frac{5}{16} e^6}.
\end{equation}
In the circular orbit limit $\left( e \rightarrow 0 \right)$, the spin frequency matches the orbital frequency, with a ratio $P_{\mathrm{PSR}}/P_{\mathrm{orb}} \rightarrow 1$. As $e \rightarrow 1$, this ratio approaches zero. For modest eccentricities the ratio is of order unity; here we consider such cases of rotation rate as characteristic ``slow'' rotators\footnote{Venus is an example of a terrestrial-sized slow rotator in our own Solar System, with a rotation period $\approx\!116$ Earth days and an orbital period of $\approx\!224$ Earth days, giving a spin-orbit rate ratio in the neighborhood of 2:1 \citep{Venus2013}.}. For higher eccentricities the ratio decreases precipitously, but only reaches periods as short as an Earth day under the MFA for $e > 0.99$ (Figure~\ref{fig:PSR_ratio}). Therefore, for all eccentricities modeled in this work, we use an Earth-like rotation period as characteristically ``fast'' rotation for comparison.

\begin{deluxetable}{ccccccc}
\tabletypesize{\footnotesize}
\tablewidth{0pt}
\tablecaption{Orbital properties for pseudo-synchronously rotating planets with Earth-like mean instellation.  The bold entries correspond to the scenarios explicitly modeled in this work.\label{table:orbital_properties}}
\tablehead{\colhead{$e$} &
           \colhead{$a$ (AU)} &
           \colhead{$P_{\mathrm{orb}}$ (days)} &
           \multicolumn{2}{c}{$P_{\mathrm{rot}}$} &
           \multicolumn{2}{c}{$F_\star$ (W m$^{-2}$)} \\
           \colhead{} &
           \colhead{} &
           \colhead{} &
           \colhead{(days)} &
           \colhead{$\left(P_{\mathrm{orb}}\right)$} &
           \colhead{Min.} &
           \colhead{Max.}
          }
\startdata
\bf{0.3} & \bf{1.024} & \bf{378.35} & \bf{242.98} & \bf{0.642} & \bf{768} & \bf{2648} \\
0.4 & 1.045 & 389.88 & 190.65 & 0.489 & 636 & 3462 \\
0.5 & 1.075 & 406.81 & 145.01 & 0.356 & 523 & 4711 \\
\bf{0.6} & \bf{1.118} & \bf{431.74} & \bf{105.89} & \bf{0.245} & \bf{425} & \bf{6800} \\
0.7 & 1.183 & 470.11 & 72.30  & 0.154 & 336 & 10791 \\
0.8 & 1.291 & 535.70 & 43.45  & 0.081 & 252 & 20400 \\
0.9 & 1.515 & 680.78 & 18.96  & 0.028 & 164 & 59281 \\
\enddata
\end{deluxetable}

\begin{figure}[htb!]
\begin{center}
   \begin{tabular}{cc}
     \includegraphics[width=8.5cm]{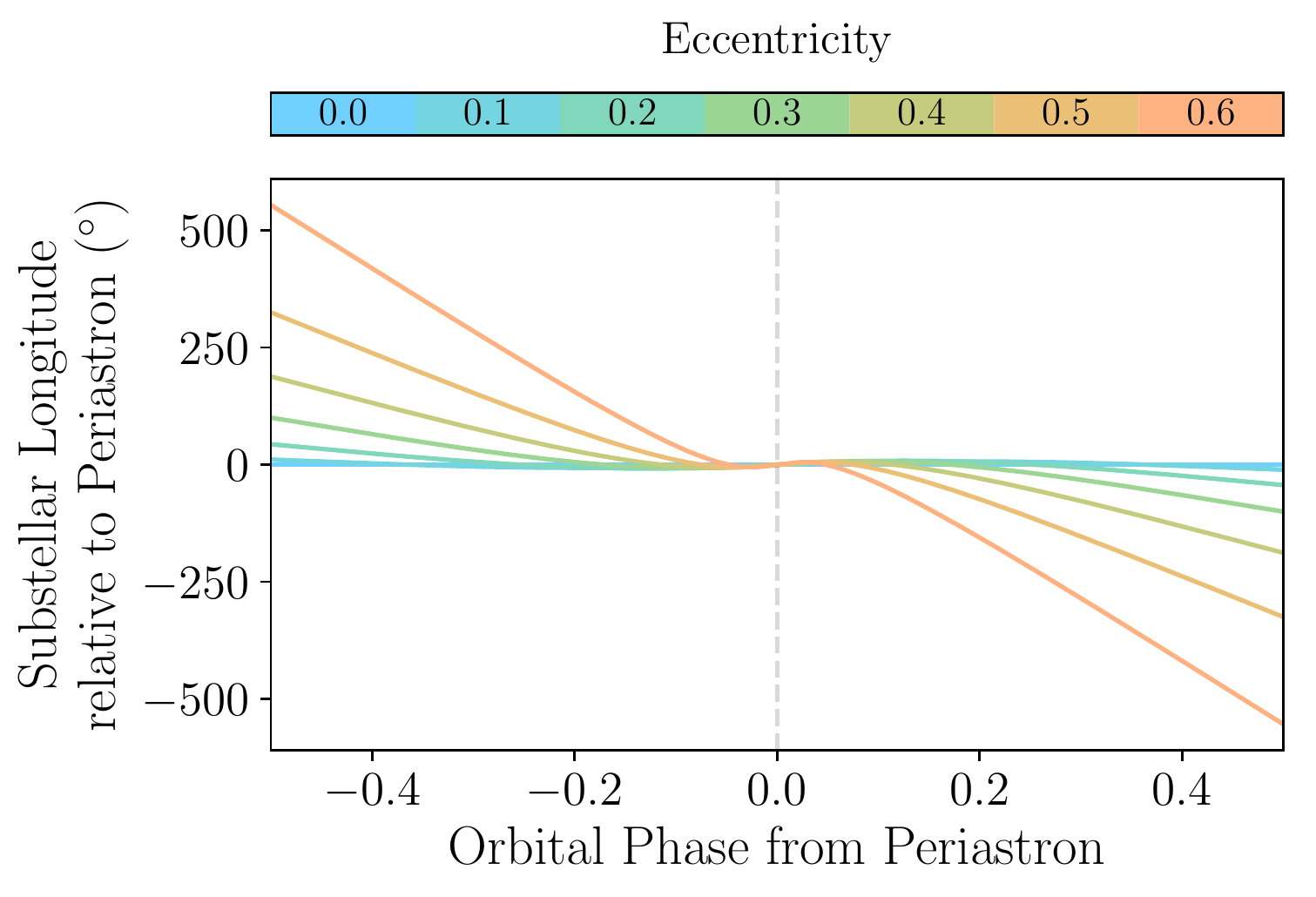} \\
     \includegraphics[width=8.5cm]{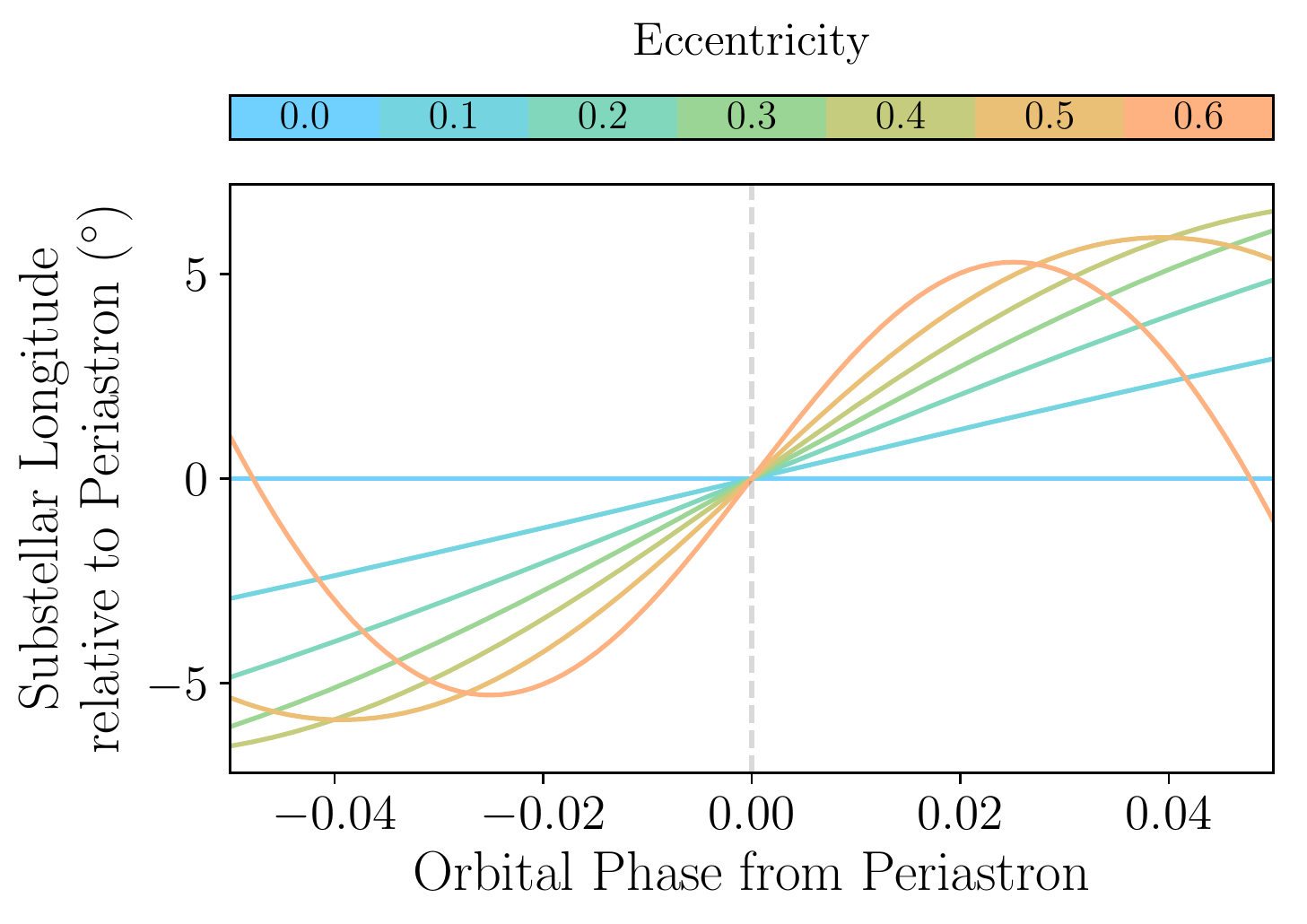} \\
   \end{tabular}
\caption{Eccentric orbits have periodically varying rates of change in the planet's orbital anomaly, with a maximum rate at periastron. For slow enough rotation (e.g. pseudo-synchronization), there exists a region of time around periastron where the instantaneous change in anomaly exceeds the rotation rate, causing the sub-stellar point on the planet's surface to move eastward rather than westward. This region is broad and the effect minor for nearly circular orbits, but narrow and increasingly intense at higher eccentricity. The plots show the movement of the sub-stellar longitude relative to its periastron position, both for one full orbit (top) as well as for a small region of orbital phase space around periastron (bottom), to show the effects at high eccentricity.}
\label{fig:double_sunrise}
\end{center}
\end{figure}

The sub-stellar longitude for a planet with zero obliquity is given by
\begin{equation}
    \phi_\star\!\left(t\right) = \phi_\star\!\left(t_0\right) - \omega_\mathrm{rot}\left(t-t_0\right) + \left[\nu\!\left(t\right)-\nu\!\left(t_0\right)\right]
\end{equation}
where $\phi_\star$ is the sub-stellar longitude, initialized to $\phi_\star\!\left(t_0\right)$ at time $t_0$, $\omega_\mathrm{rot}$ is the angular rate of rotation, and $\nu$ is the true anomaly of the orbit. At high eccentricities, the sub-stellar point on the planet's surface exhibits a reversal of its direction of motion around periastron. This effect is due to the relation between the planet's rotation rate and the variable rate of change in true anomaly, and is also referred to as ``optical libration''. Works such as \citet{Selsis2013} and \citet{bol16} have demonstrated that, for models with planets on eccentric orbits and spin-orbit synchronization, the sub-stellar point will librate over an orbit, preventing perpetual day and night sides. Figure~\ref{fig:double_sunrise} shows this effect on the sub-stellar longitude as a function of eccentricity for pseudo-synchronous rotation. The rate of change in anomaly $\dot{\nu}$ at periastron, relative to the mean motion $n \equiv 2\pi/P_\mathrm{orb}$, is
\begin{equation}\label{eq:anomaly_rate}
\frac{\dot{\nu}_{\mathrm{peri}}}{n} = \sqrt{\frac{1+e}{\left(1-e\right)^3}}.
\end{equation}
This evaluates to 5 for $e=0.6$, compared with roughly 4 for $\omega_\mathrm{rot}/n$ under pseudo-synchronous rotation. This transient increase in the rate of change in true anomaly causes it to exceed the rotation rate around periastron, during which time the sub-stellar point reverses direction from its otherwise westward motion\footnote{This also happens, for example, on Mercury, whose moderate orbital eccentricity of $\approx\!0.2$ and 3:2 spin-orbit rate ratio causes this effect.}. While this effect can greatly expand the range of illuminated longitudes at high eccentricities for $\omega_\mathrm{rot}/n=1$, for pseudo-synchronization the range of this motion is limited to a few degrees around the sub-stellar longitude around periastron, remaining close to an approximation of tidal locking. The resulting effect is that a significant fraction of the period of greatest instellation is spent concentrated on a confined set of longitudes, rather than distributed nearly uniformly as in the fast rotation case. This concentration of stellar heating drives strong thermal and dynamical responses in the atmosphere, whose observable effects we detail in \S \ref{sec:results}.

\section{Model Details}\label{sec:model}

The National Center for Atmospheric Research (NCAR) Community Atmospheric Model (CAM) is a global model designed to simulate Earth's atmosphere \citep{nea10}, and is the atmospheric component of the fully coupled Community Earth System Model (CESM).  The physical evolution of the atmosphere in CAM is represented by the Navier-Stokes equations under the approximations of vertical hydrostatic equilibrium and a shallow aspect ratio of the flow, which together constitute the primitive equations. The primitive equations are implemented in a finite volume dynamical core that uses  horizontally Eulerian and vertically Lagrangian discretization to account for the grid-scale motions of dry air \citep{lin96,lin97}, with additional conservation equations for water.  The dynamical core conserves mass, momentum, and total energy with numerics that ensure physical tracers (e.g. water vapor) remain non-negative at each time step \citep{nea10}.  A suite of sophisticated algorithms are used to represent the net effect of subgrid-scale processes on the grid-scale variables; the subset of these parameterizations that seem most relevant to observables are described in the remainder of this section.

Here we use an adaptation of the ExoCAM\footnote{\url{https://github.com/storyofthewolf/ExoCAM}} extension of CAM 4. Short for Exoplanet CAM, ExoCAM is designed for studies of both exoplanets and deep-time paleoclimates on Earth, with particular attention to expanding the valid ranges of stellar forcing, atmospheric partial pressures of greenhouse gases (H$_2$O, CO$_2$, and CH$_4$) \citep{wol13,wol14}, and planetary rotation rate. In particular, ExoCAM improves one of the numerical solvers in the deep convection scheme, using the more robust approach of \citep[][courtesy C. A. Shields]{Brent1973}. This scheme attempts to represent the effect of the total vertical mass flux in ensembles of cumulus clouds on grid-scale atmospheric temperatures and humidities. ExoCAM also includes improved substepping in the dynamical core, by applying fractional physics tendencies across the dynamical substeps instead of only at the beginning of the timestep \citep{Bardeen2017}.  This feature improves model numerical stability for slow rotators and high incident solar fluxes. ExoCAM also extended to higher temperatures the absorption coefficients for the correlated-$k$ radiative transfer scheme.  These features allow ExoCAM to operate under warm and moist greenhouse conditions, with mean surface temperatures $\lesssim 365$ K and water vapor partial pressures $\lesssim 0.2$ bar. 

Our version of the model is similar to the configuration described in \citet{Kopparapu2017}. The horizontal resolution is $4^\circ\!\times\!5^\circ$, with 40 vertical levels ranging from a global mean surface pressure around 1 bar to a minimum pressure of $10^{-3}$ bar. We employ a commonly used aquaplanet configuration for simplicity: we fix the planetary radius and surface gravity to Earth values, and have flat topography covered by a uniform-depth slab ocean \citep{Bitz2012}. Most of our simulations use an ocean depth of 50 meters, but we also ran with ocean depths of 10 meters at $e=0.3$ to study how our results might change with ocean depth (see \S \ref{sec:discussion}). Ocean heat transport is neglected, but the model accounts for sea ice formation via the CICE model from \citet{hun08}. The water ocean albedos in both the visible and near-IR are tuned to 0.06 for direct and 0.07 for diffuse reflection, matching the values established in \citet{shi13}.

Bulk microphysical processes of condensation, precipitation, and evaporation follow the methods of \citet{Rasch1998}. Deep convection is treated using the method of \citet{Zhang1995}, that has been further updated to include convective momentum transport and dilute entraining plumes \citep{Raymond1986,Raymond1992}. A separate convective treatment is employed for shallow adjustments following \citet{Hack1994}. In each grid cell, changing water-vapor amounts are self-consistently handled, with the total parcel mass determined from advection, convection, turbulent mixing, and large-scale stable condensation and evaporation tendencies. Virtual temperature corrections account for variations in density and the specific heat of moist air. Cloud fractions are calculated separately for for marine stratus, convective clouds, and layered clouds. Cloud liquid droplet radii are assumed to be 14 mm everywhere in the model. Ice cloud particle effective radii follow a temperature-dependent parameterization and can vary in size from a few tenths of microns to a few tenths of millimeters.

The radiative transfer code uses the correlated-k absorption coefficients derived from the HITRAN 2012 spectral database using the HELIOS-K open source spectral sorting program \citep{Kopparapu2017}, utilizing the standard two-stream approach from \citet{Toon1989}. The spectral binning (Table \ref{table:spectral_bands}) encompasses the solar spectrum, spanning 0.2--12.2 $\mu$m (Bands 1--35 in our numbering scheme), as well as the planet's thermal emission in the range 2.5--1000 $\mu$m (Bands 24--42). The spectral intervals used to calculate the relevant stellar and planetary fluxes are often referred to as the ``shortwave'' and ``longwave'', which are not strict divisions since the wavelength ranges overlap in ExoCAM. Accordingly, we use the ExoCAM solar spectrum for our stellar fluxes. Our initial atmosphere is assumed to be purely N$_2$, with the only greenhouse gas being H$_2$O drawn from the surface ocean; the water vapor continuum is treated using the formalism of \citet{Paynter2011}. The radiative effects of cloud overlap are treated using the Monte Carlo Independent Column Approximation (MCICA), assuming maximum-random overlap \citep{Pincus2003}.

\begin{deluxetable}{ccccc}
\tabletypesize{\small}
\tablewidth{0pt}
\tablecaption{Wavelength Ranges of Model Spectral Bands\label{table:spectral_bands}}
\tablehead{\colhead{Band} &
           \colhead{$\lambda_{\mathrm{start}}$} &
           \colhead{$\lambda_{\mathrm{mid}}$} &
           \colhead{$\lambda_{\mathrm{end}}$} &
           \colhead{Brightness Temp.} \\
           \colhead{} &
           \multicolumn{3}{c}{($\mu$m)} &
           \colhead{(K)}
          }
\startdata
1  & 0.200 & 0.227 & 0.263 & 11011--14489 \\
2  & 0.263 & 0.299 & 0.345 & 8404--11011  \\
3  & 0.345 & 0.387 & 0.442 & 6563--8404   \\
4  & 0.442 & 0.477 & 0.518 & 5593--6563   \\
5  & 0.518 & 0.567 & 0.625 & 4636--5593   \\
6  & 0.625 & 0.642 & 0.660 & 4390--4636   \\
7  & 0.660 & 0.676 & 0.692 & 4187--4390   \\
8  & 0.692 & 0.717 & 0.743 & 3898--4187   \\
9  & 0.743 & 0.760 & 0.778 & 3724--3898   \\
10 & 0.778 & 0.811 & 0.847 & 3419--3724   \\
11 & 0.847 & 0.877 & 0.909 & 3188--3419   \\
12 & 0.909 & 0.952 & 1.00  & 2898--3188   \\
13 & 1.00  & 1.05  & 1.10  & 2637--2898   \\
14 & 1.10  & 1.17  & 1.24  & 2333--2637   \\
15 & 1.24  & 1.27  & 1.30  & 2231--2333   \\
16 & 1.30  & 1.36  & 1.43  & 2028--2231   \\
17 & 1.43  & 1.47  & 1.50  & 1927--2028   \\
18 & 1.50  & 1.56  & 1.63  & 1782--1927   \\
19 & 1.63  & 1.69  & 1.77  & 1637--1782   \\
20 & 1.77  & 1.85  & 1.94  & 1492--1637   \\
21 & 1.94  & 1.99  & 2.04  & 1420--1492   \\
22 & 2.04  & 2.09  & 2.15  & 1347--1420   \\
23 & 2.15  & 2.31  & 2.50  & 1159--1347   \\
24 & 2.50  & 2.76  & 3.08  & 942--1159    \\
25 & 3.08  & 3.42  & 3.84  & 753--942     \\
26 & 3.84  & 4.02  & 4.20  & 690--753     \\
27 & 4.20  & 4.37  & 4.55  & 638--690     \\
28 & 4.55  & 4.67  & 4.81  & 603--638     \\
29 & 4.81  & 5.15  & 5.56  & 522--603     \\
30 & 5.56  & 6.10  & 6.76  & 429--522     \\
31 & 6.76  & 6.97  & 7.19  & 403--429     \\
32 & 7.19  & 7.78  & 8.47  & 342--403     \\
33 & 8.47  & 8.77  & 9.09  & 319--342     \\
34 & 9.09  & 9.62  & 10.20 & 284--319     \\
35 & 10.20 & 11.11 & 12.20 & 238--284     \\
36 & 12.20 & 13.16 & 14.29 & 203--238     \\
37 & 14.29 & 15.04 & 15.87 & 183--203     \\
38 & 15.87 & 17.70 & 20.00 & 145--183     \\
39 & 20.00 & 21.62 & 23.53 & 123--145     \\
40 & 23.53 & 25.81 & 28.57 & 101--123     \\
41 & 28.57 & 36.36 & 50.00 & 58--101      \\
42 & 50.00 & 525   & 1000  & 3--58        \\
\enddata
\tablecomments{Bands 30--41 were used in our analysis. We include the entire model spectral output for completion.}
\end{deluxetable}

\subsection{Orbital Configuration}\label{eccmods}
We ran models at orbital eccentricities of 0.3 and 0.6, modifying the orbital calculations used in the standard ExoCAM code to ensure greater accuracy at high eccentricity. For high eccentricities, calculating the orbital position is crucial for correctly modeling the sub-stellar position and time-dependent instellation. By default, CESM calculates the true longitude (corresponding to the true anomaly for exoplanets) of the Earth using an approximation given by a third-order polynomial,
\begin{align}
\begin{split}
  \lambda = M &+ \left(2e - \frac{e^3}{4} \right) \sin\!\left(M + \varpi \right)\\
  &+ \frac{5e^2}{4} \sin\!\left[2\left(M + \varpi \right)\right]\\
  &+ \frac{13e^3}{12} \sin\!\left[3\left(M + \varpi \right)\right],
\end{split}
\end{align}
where $M$ is the mean anomaly, $e$ is the orbital eccentricity, and $\varpi$ is the longitude of periastron. This approximation is valid to $\lesssim 0.3\%$ for eccentricities up to 0.1. However, the approximation rapidly diverges from the exact result, with the error reaching $\approx 12\%$ at $e=0.5$, and $\approx 83\%$ at $e=0.95$. We replace the approximation with a simple numerical method that iteratively solves Kepler's equation,
\begin{equation}
M = E - e \sin E,
\end{equation}
where $E$ is the eccentric anomaly.

\subsection{Observable Properties}\label{sec:model:observables}

\begin{figure}[htb!]
\begin{center}
    \includegraphics[width=8.5cm]{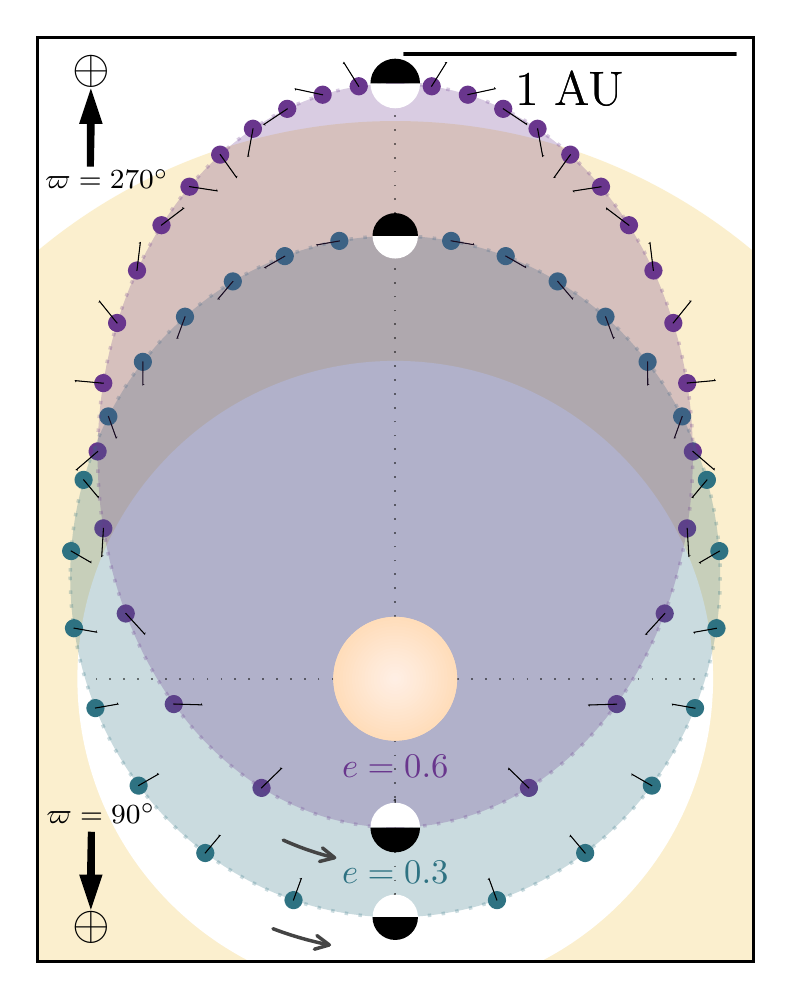}
\caption{We consider two orbital eccentricities (0.3 and 0.6), each from two possible edge-on viewing geometries. At $\varpi=90^\circ$, we see the planet's night side (anti-stellar hemisphere) during its periastron passage, and the day side (sub-stellar hemisphere) during apastron. At $\varpi=270^\circ$, the day side viewing is aligned with periastron, and the night side during apastron. The Habitable Zone limits from \citet{kop14} are plotted as a golden annulus for Earth-mass planets, with the ``Runaway Greenhouse'' and ``Maximum Greenhouse'' limits setting the inner and outer radii, respectively. The colored dots represent points equally-spaced in time, at intervals of $\approx\!13.5$ days; their poles show the rotation through one orbit (from apastron to apastron) for pseudo-synchronous rotation.}
\label{fig:orbital_configuration}
\end{center}
\end{figure}

We now describe how we simulate the radiation that would be observed by a telescope at a location distant from the planet and star of interest. Exoplanetary systems can in general have any possible orientation with respect to the observer. For our full-phase analyses we restrict ourselves to 2 lines of sight, both edge-on with respect to the orbital plane: one along the periastron-star line and another along the apastron-star line (Figure \ref{fig:orbital_configuration}). This amounts to varying the longitude of periastron while keeping other orientation parameters fixed. In the first case (corresponding to an inclination $i=90^\circ$, longitude of periastron $\varpi = 90^\circ$), one would observe the night side of the planet during periastron, when the planet would transit its host star; then, one would be able to see the day side during apastron, when the planet passes through its secondary eclipse. In the second case ($i=90^\circ$, $\varpi = 270^\circ$), the day side is visible during periastron (or more precisely, just before and after eclipse), and the night side (transit) at apastron. In addition to these two cases, we also generate theoretical eclipse depths for the full possible range of longitudes of periastron, in order to show the variation in day-side emission with viewing geometry.

To generate theoretical light curves, we take the outgoing radiation maps in each model band and calculate the expected observable flux for a given viewing geometry. We use the ExoCAM solar spectrum for the model host star to then express our expected brightnesses in units of the thermal stellar flux. To compare with a realistic set of observations in the infrared, we adopt the wide filter profiles for the Mid-Infrared Imager \citep[MIRI, see][]{miri_doc} on the James Webb Space Telescope \citep[JWST, see][]{rie15,wri15,gla15}. These filters span a range of approximately 5--30 $\mu$m. Convolving these filter profiles with the model bands, we generate emission maps for each MIRI filter\footnote{These bands are adopted as a template for mid-infrared photometry. The precision required to discern the variations in this study will be outside the scope of JWST itself; for a discussion of the potential for future observations, see \S \ref{sec:discussion}.}. Given an emission map $M_\lambda\!\left(\phi,\theta,t\right)$ at a specific wavelength, we solve for the corresponding planet-star flux contrast via
\begin{equation}\label{eq:contrast}
\bar{F}\!\left(t\right) = \frac{1}{\pi} \left(\frac{R_\mathrm{p}}{R_\star}\right)^2 \frac{\iiint w M_\lambda V \,d\lambda\,d\theta\,d\phi}{\int w F_{\lambda,\star} \,d\lambda}
\end{equation}
where $w=w\!\left(\lambda\right)$ is the weighted response of the instrumental bandpass at $\lambda$, $V=V\!\left(\phi,\theta,t\right)$ is the component of the normal vectors of the planet grid cells along the line of sight, and $F_{\lambda,\star}$ is the disk-integrated stellar flux at $\lambda$. For a sub-observer point given by longitude $\phi_\mathrm{obs} = \phi_\star\!\left(t_\mathrm{ecl}\right) - \omega_\mathrm{rot}\left(t-t_\mathrm{ecl}\right)$ for rotation rate $\omega_{\mathrm{rot}}$ and a reference eclipse midpoint time satisfying $\nu\!\left(t_\mathrm{ecl}\right) = \frac{3}{2}\pi - \varpi$, and latitude $\theta_\mathrm{obs} = 0$,
\begin{equation}\label{eq:visibility}
V = 
\begin{cases}
\cos\left(\phi-\phi_\mathrm{obs}\right) \cos\theta, &\cos\left(\phi-\phi_\mathrm{obs}\right) \leq \pi/2 \\
0, &\cos\left(\phi-\phi_\mathrm{obs}\right) > \pi/2
\end{cases}
\end{equation}
Once we have a full orbit of predicted photometry, we select the contrasts during eclipse as our predicted eclipse depths.

Each model is run for 25 orbital periods; this value was chosen as the longest time needed (across all of our integrations) for the range of global mean, time mean surface temperatures for each of a span of 10 orbits to be within 1\% of the global mean, time mean temperature averaged over the same 10-year span. Within the final 10 years, the model also remains within 0.25 W m$^{-2}$ of global, annual mean radiative balance at the top of the atmosphere. Our spin-up time of 15 years is similar to other slab-ocean GCM simulations, matching for example that of \citet{Chiang2005}, who also use 50-meter ocean mixed layer depths. The model photometry we present accordingly uses statistics calculated for the final 10 orbits of each simulation.

\section{Results}\label{sec:results}
\subsection{Internal Results}\label{sec:results:internal}

In this section we describe the simulated climate of our four hypothetical planets, then illustrate a key difficulty in interpreting observations of such planets: temperatures in the upper troposphere near the emission level vary because of both day-night contrast and orbital periodicity, with the amplitude and phase of these variations being highly sensitive to orbital eccentricity.

\subsubsection{Surface Conditions}\label{sec:results:internal:surface}

\begin{figure*}
\begin{center}
   \begin{tabular}{cc}
     \includegraphics[width=8.5cm]{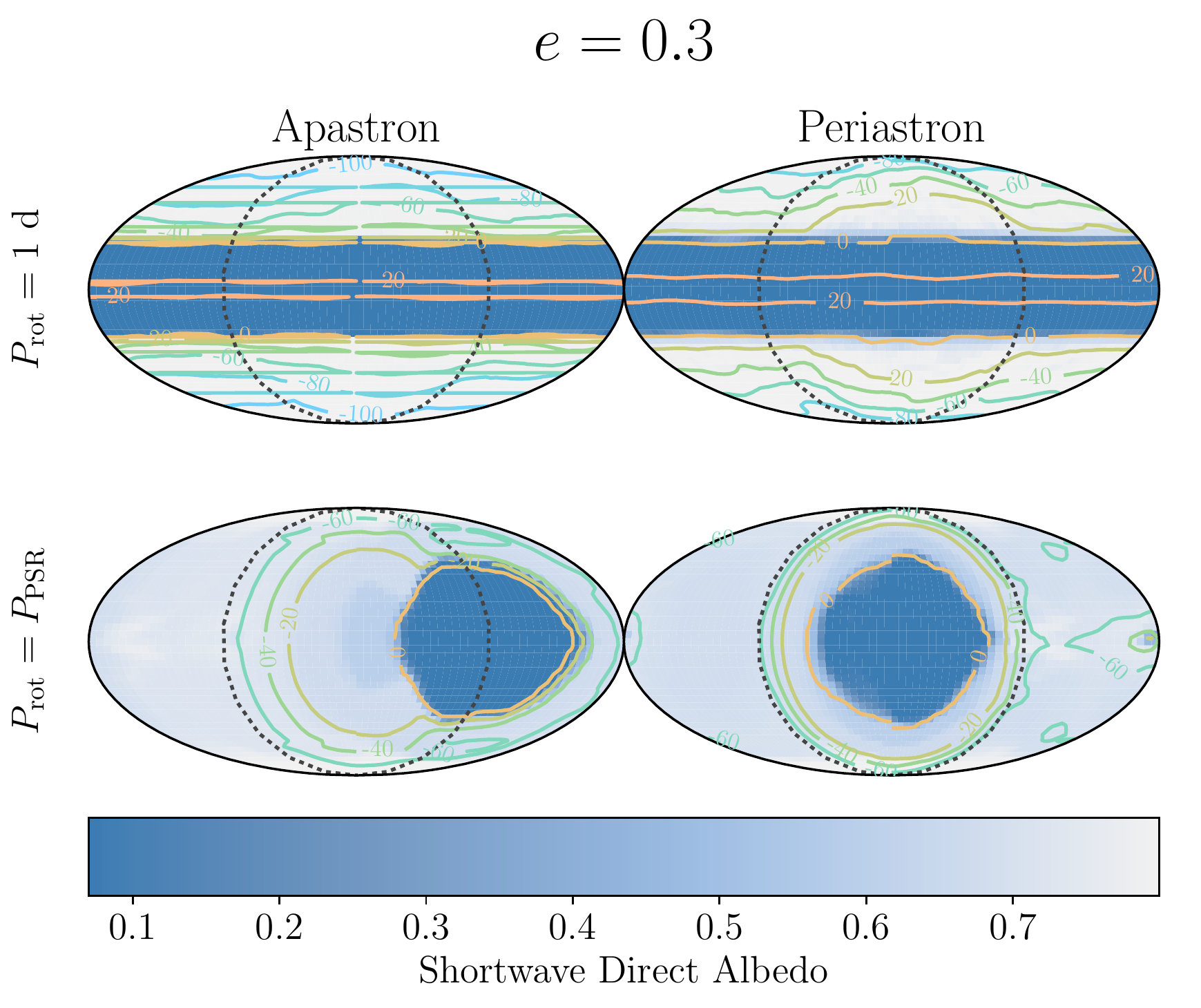} &
     \includegraphics[width=8.5cm]{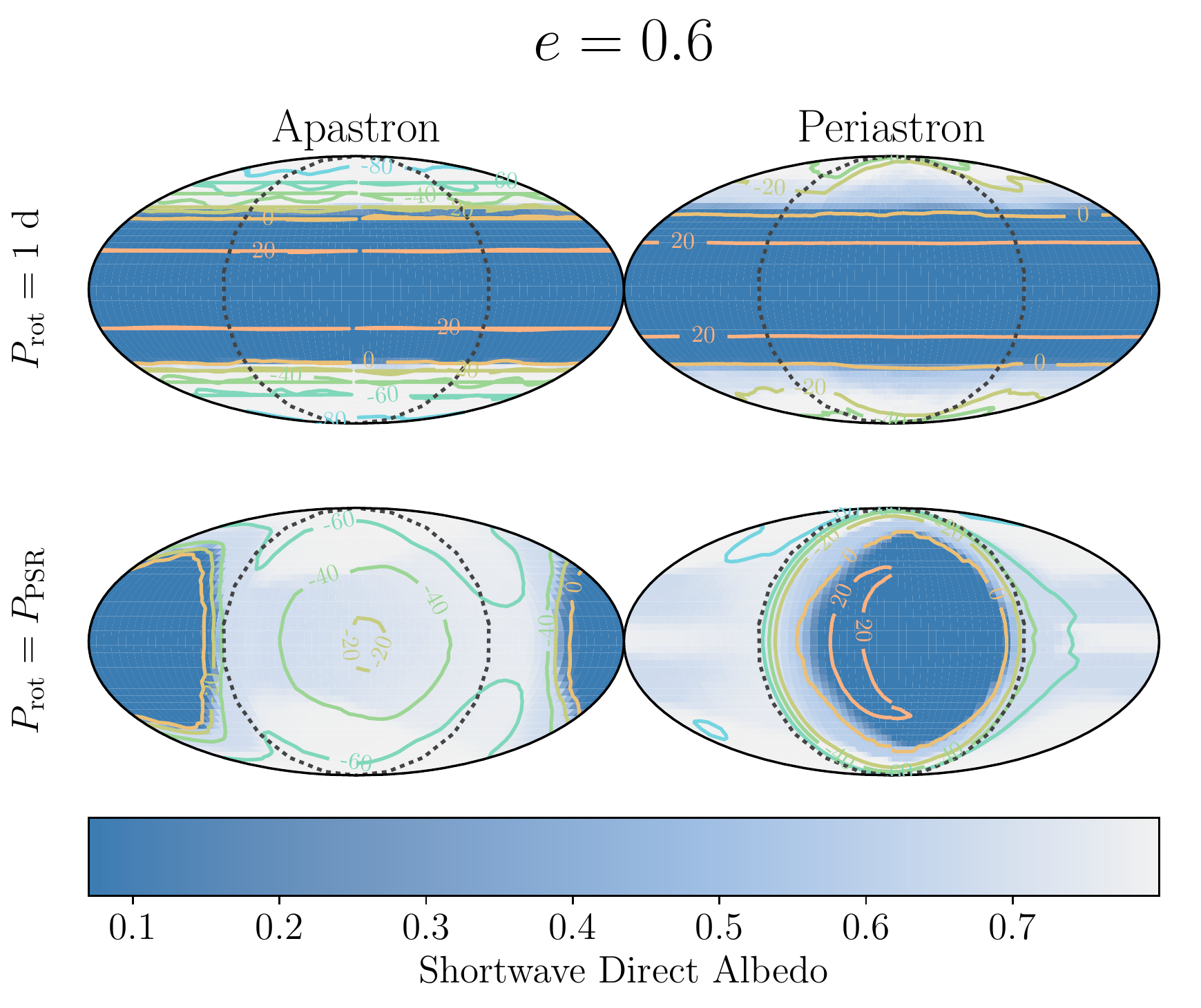} \\
   \end{tabular}
\caption{Global maps of the diffuse shortwave albedo at the surface at the extreme points of each orbit, shown for fast rotation (Earth-like, in the upper row) and the much slower pseudo-synchronous rotation, defined in Equation \ref{eq:PSR}, in the lower), at orbital eccentricities of 0.3 and 0.6. The dark dotted lines delineate the star-facing hemisphere, which is centered in each plot. Colored contours denote surface temperature in Celsius. The nominal albedo for liquid water is taken to be 0.06--0.07 (see \S \ref{sec:model} for details), and the limiting albedo for thick ice cover is 0.8.}
\label{fig:ALBEDO_globes}
\end{center}
\end{figure*}

Rotation rate strongly influences horizontal temperature gradients, as expected, with fast rotation confining warm air near the equator and producing strong eastward winds that homogenize energy content in longitude.  This is clearly seen in the distributions of surface temperature and surface albedo, with the latter indicating the regions covered by sea ice on these aquaplanets (Figure~\ref{fig:ALBEDO_globes} and Table \ref{table:surface_temperatures}). Temperature and albedo contrasts are primarily latitudinal for Earth-like rotation, and are accompanied by a Hadley circulation with a rising branch centered on the equator. For slow, pseudo-synchronous rotation, longitudinal contrasts are just as strong as latitudinal ones, with warm, ice-free oceans centered on the sub-stellar points at periastron and ice-covered regions extending over the night side and polar regions. The slow rotator circulation patterns are also much broader in latitude than their fast counterparts, in agreement with the circulation patterns seen in previous simulations of aquaplanets at different rotation rates, e.g. Figure 6 in \citet{mer10}. At apastron, the effects of ocean thermal inertia and deviations from strict synchronous rotation combine to shift the warm, ice-free region away from the sub-stellar point.  For $e$ = 0.3 and slow rotation, the ice-free region straddles the day-night line at apastron; for $e$ = 0.6 and slow rotation the ice-free region is actually on the night side of the planet at apastron while the day side is ice-covered and has a secondary temperature maximum of about $-20^\circ$ C at the sub-stellar point.

\begin{deluxetable*}{cccccccc}
\tabletypesize{\small}
\tablewidth{0pt}
\tablecaption{Global Ranges and Means of Surface Temperature (K)\label{table:surface_temperatures}}
\tablehead{\colhead{} &
		   \colhead{} &
           \multicolumn{3}{c}{$\oplus$} &
           \multicolumn{3}{c}{PSR} \\
           \colhead{e} &
           \colhead{Time} &
           \colhead{Range} &
           \colhead{Mean} &
           \colhead{Ocean Mean\tablenotemark{a}} &
           \colhead{Range} &
           \colhead{Mean} &
           \colhead{Ocean Mean\tablenotemark{a}}
          }
\startdata
\multirow{2}{*}{0.3} & Apo & 155--295 & 220 & 288 & 198--286 & 228 & 280 \\
& Peri & 170--295 & 241 & 289 & 195--285 & 228 & 278 \\
\multirow{2}{*}{0.6} & Apo & 183--305 & 248 & 292 & 200--287 & 228 & 280 \\
& Peri & 214--305 & 271 & 293 & 191--298 & 232 & 285 \\
\enddata
\tablenotetext{a}{The ocean mean is defined as the global mean of all regions with $T>273.15$ K.}
\end{deluxetable*}

The horizontal range in surface temperature at both extremes of the orbit is larger for fast rotation than for slow, implying that rotational confinement by the atmospheric circulation is more effective at generating horizontal temperature gradients than the day-night contrast in radiative heating.  In fact, both the range and global mean of surface temperatures for the $e=0.3$, slow-rotation case are remarkably similar between apastron and periastron. \citet{Cowan2012} calculated that, for a terrestrial planet, the effective thermal relaxation times for snowball and temperate (ocean) climates are $\sim\!145$ and 343 days, respectively. Given that this scale is much longer than the analogous timescales expected for much of the atmospheres of eccentric Hot Jupiters \citep[e.g.][]{Kataria2013,dew16}, it is not surprising that surface temperature constrasts between the extremes of the orbit are quite muted for the slow rotator. The differences in apastron-to-periastron global mean temperature for the fast rotators at both eccentricities exceed $20^\circ$C, a greater but still modest contrast given the differences in instellation between the extremes of the orbits.

\subsubsection{Atmospheric Conditions}\label{sec:results:internal:atmosphere}

\begin{figure*}
\begin{center}
   \begin{tabular}{c}
     \includegraphics[width=17cm]{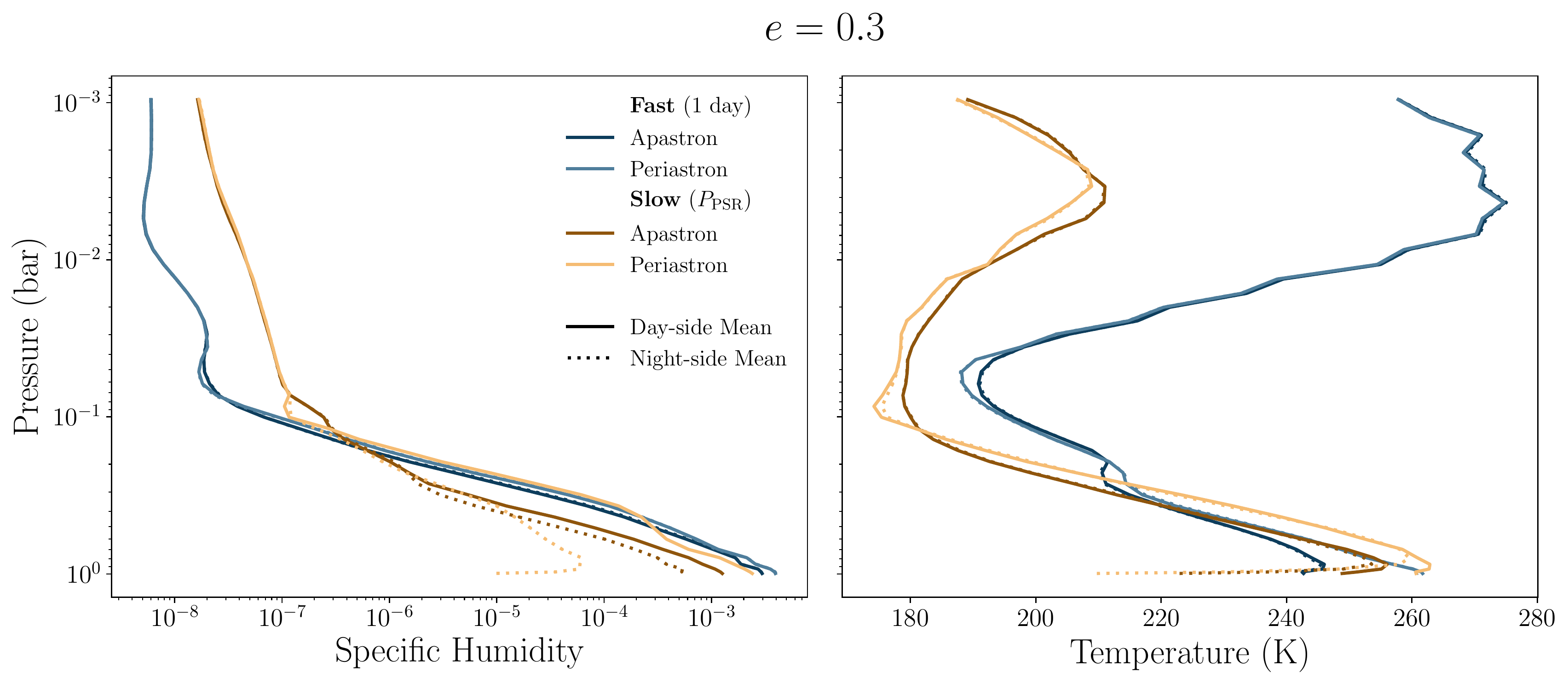} \\
     \includegraphics[width=17cm]{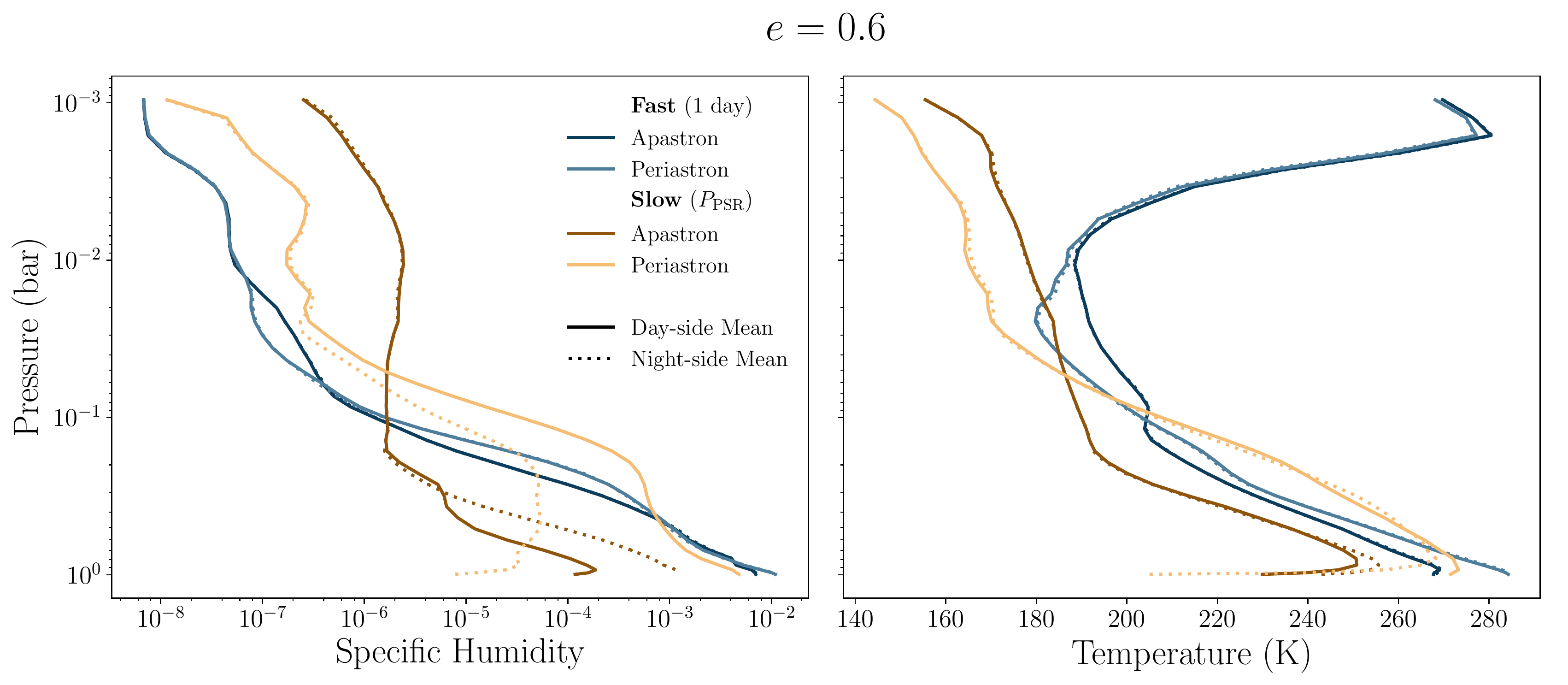} \\
   \end{tabular}
\caption{The specific humidity (fraction of atmospheric mass in H$_2$O) and temperature, both averaged over the day sides (solid lines) and night sides (dotted lines). Each color represents a rotation period: the bluer hues show fast (Earth-like) rotation, and the golden hues show slow (pseudo-synchronous) rotation. The lighter shades represent each quantity during periastron, and the darker shades each during apastron.}
\label{fig:Q-P_T-P}
\end{center}
\end{figure*}

Temperature contrasts are even more muted between the day and night side of the slow rotator in the atmosphere above the lowermost troposphere (Figure~\ref{fig:Q-P_T-P}, right column). This is expected because the Rossby deformation radius is larger than the planetary circumference at these rotation rates, allowing atmospheric circulations to rapidly homogenize temperatures throughout most of the troposphere, as in previous studies of strict synchronous rotation \citep[e.g.][]{jos97}. But in contrast with those previous studies, the shallow, near-surface temperature inversion that forms on the ice-covered side of the planet is actually on the day side of the planet at apastron (see the bottom right panel of Figure~\ref{fig:Q-P_T-P}), due to the combined effects of ocean thermal inertia and deviations from strict synchronous rotation, as discussed above. As noted in \S \ref{sec:results:internal:surface}, the thermal timescales for surface ice and water are much longer than the timescales at equivalent temperatures in the upper atmosphere, which themselves are longer than the typical Hot Jupiter timescales due to the much lower average atmospheric temperatures for the ocean planets \citep{Showman2002,Showman2010}.

\begin{figure}[htb!]
\begin{center}
   \begin{tabular}{c}
     \includegraphics[width=8.5cm]{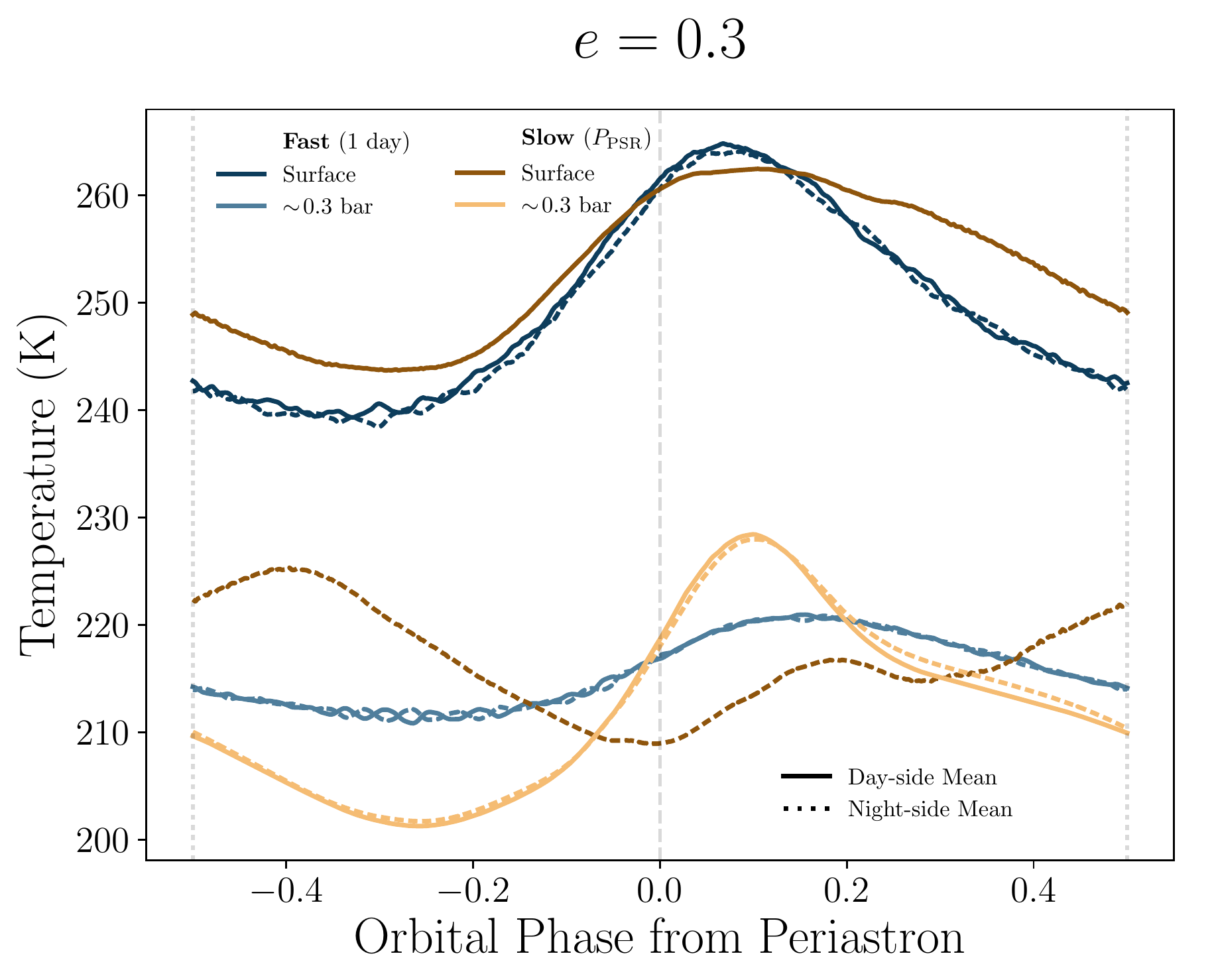} \\
     \includegraphics[width=8.5cm]{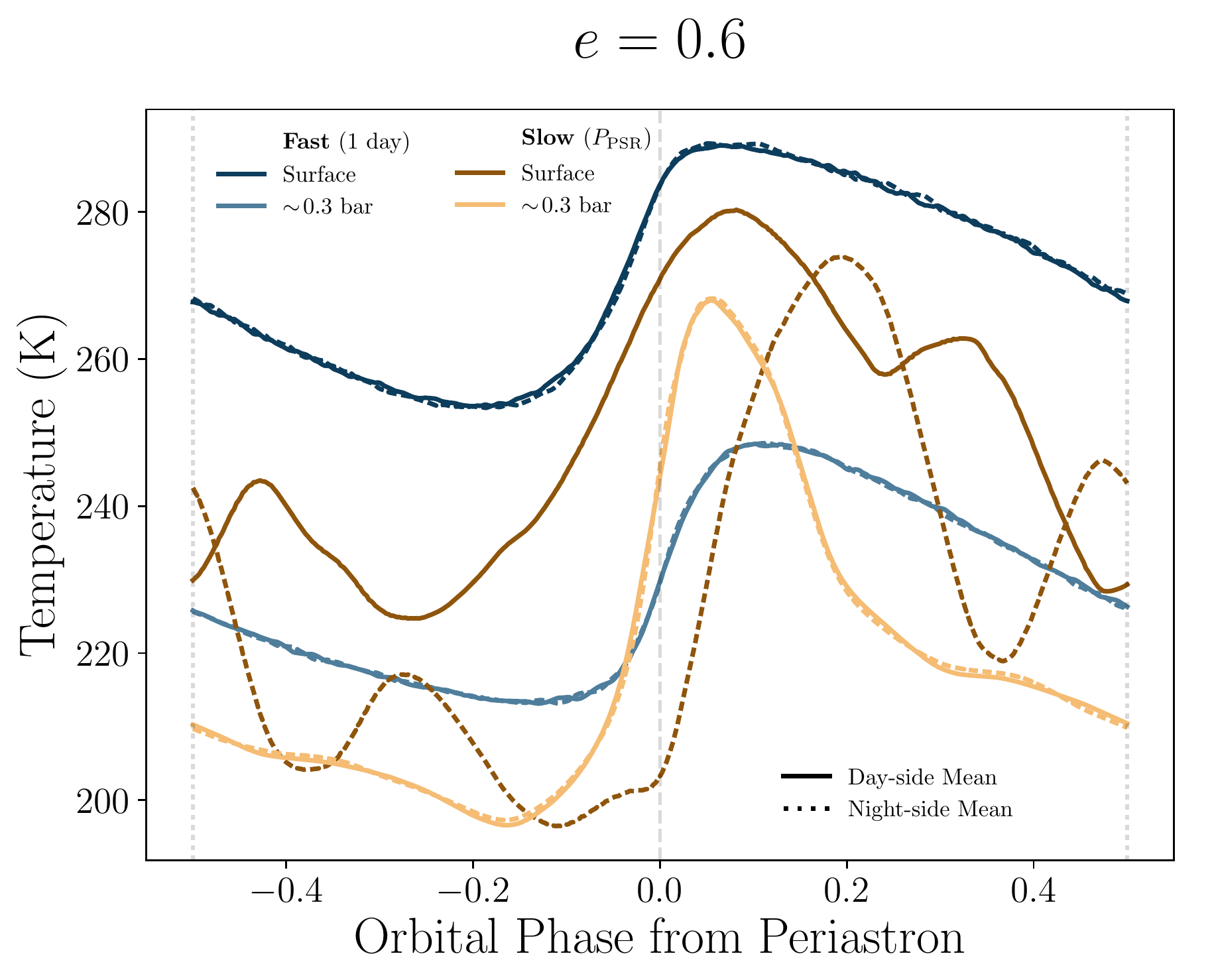} \\
   \end{tabular}
\caption{Time series of hemispherically-averaged temperatures over one orbit. Each color represents a rotation period: the bluer hues show fast (Earth-like) rotation, and the golden hues show slow (pseudo-synchronous) rotation. Lighter shades are the temperatures at the effective emission layer (taken to be centered at $\sim\!0.3$ bar, averaged over 0.2--0.4 bar), and darker shades are surface temperatures. The solid lines are the averages over the day sides and the dotted lines the averages over the night sides. At the surface, day-night contrasts in temperature are comparable in amplitude to the variations due to the orbital periodicities in instellation. However in the upper troposphere, from where outgoing emission may originate in many longwave spectral bands, the day-night contrasts are effectively negligible.}
\label{fig:uppertrop_T}
\end{center}
\end{figure}

To illustrate how these effects might complicate interpretation of observations, we plot time series, over an orbit, of the day- and night-side temperatures averaged over an atmospheric layer near the emission level, which we estimate to be $\sim\!0.3$ bar (Figure~\ref{fig:uppertrop_T}).  In an optically thick atmosphere, radiative emission might come from this upper-tropospheric layer, making its temperature more relevant to observations than surface conditions.  The behavior of the fast rotator is easy to understand, with day- and night-side temperatures being nearly equal, and peak temperatures reached shortly after periastron for both eccentricities.  For the slow rotator, the day-night contrast in upper-level temperature is also very small, and the orbital variations in emission from this level are due almost entirely to changes in planet-star distance.  It would thus be difficult to distinguish a fast rotator from a slow rotator given only emission from this upper-tropospheric level.  In contrast, surface temperature exhibits a large contrast between day side and night side on the slow rotators.  Night-side surface temperature has the same range as 0.3 bar temperature but with the opposite phase for $e$ = 0.3; for $e$ = 0.6, night-side surface temperatures for the slow rotator show several peaks over an orbit, corresponding to the roughly four rotations that occur over an orbit at this eccentricity (Fig.~\ref{fig:PSR_ratio}).  Thus, inferences about rotation rate would be most easily made from the day-night contrast in surface temperature, but it is possible that emission will come from a much higher altitude.

The troposphere of the slow (pseudo-synchronous) rotator exhibits a much wider longitudinal contrast in water content than in temperature (Figure~\ref{fig:Q-P_T-P}, left column). As we move from the surface to the upper troposphere, the slow rotator retains a much moister troposphere on the day side at periastron, even though temperatures are nearly equal above the lower-tropospheric inversion layer.  This is also true at apastron for $e$ = 0.3, but for $e$ = 0.6 at apastron the humidity is higher over the open ocean on the night side and lower on the day side over the ice-covered surface.  Considering the slowly rotating case further, we notice that the strong wet-dry difference extends to $\sim\!0.1$ bar at periastron for $e$ = 0.3, compared with a weaker wet-dry difference extending to $\sim\!0.25$ bar at apastron. This reflects the large warming and deepening of the troposphere over the sub-stellar point when the radiative forcing is strongest at periastron.  The day-night humidity contrast seen on the slow rotator is even more pronounced at higher eccentricity, reaching to $\sim\!0.03$ bar.  The upper troposphere at apastron is warmer and more humid at the sub-stellar point than at the anti-stellar point, even though the converse is true at the surface and in the lower troposphere, showing the complexity introduced to day-night contrasts by pseudo-synchronous rotation.

\begin{figure*}
\begin{center}
   \begin{tabular}{cc}
     \includegraphics[width=8.5cm]{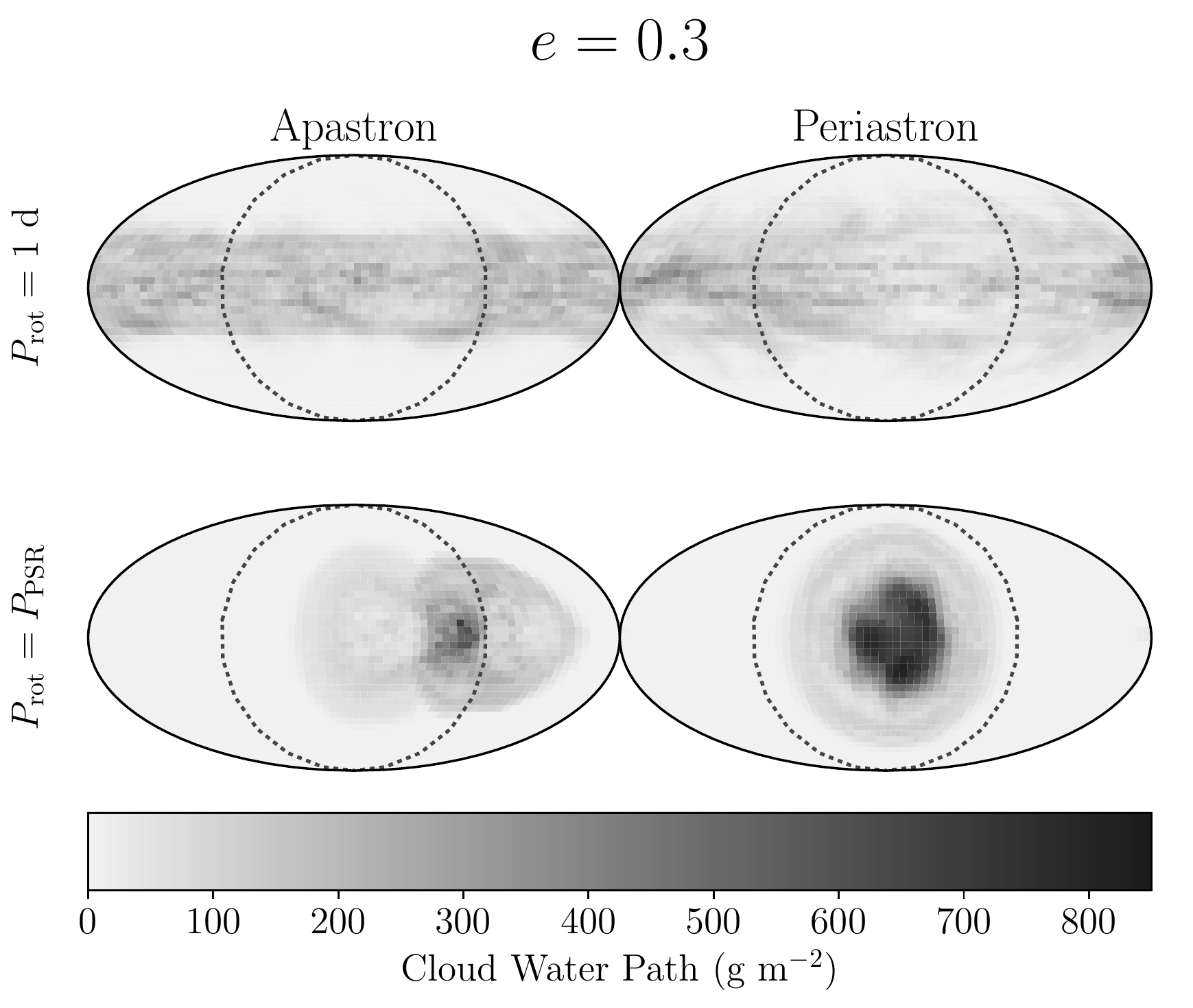} &
     \includegraphics[width=8.5cm]{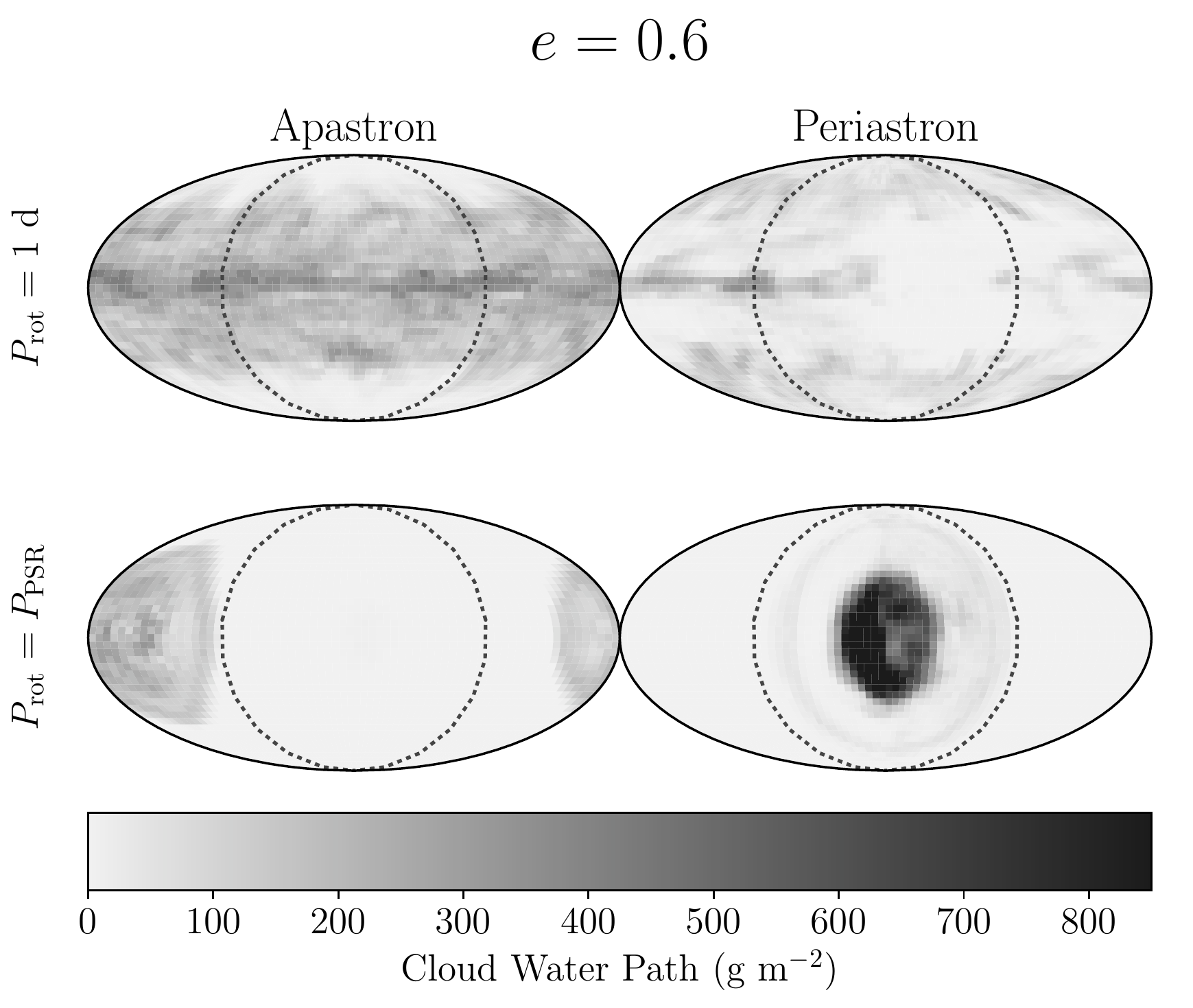} \\
   \end{tabular}
\caption{Global maps of the vertically integrated condensed cloud water at the extreme points of each orbit, shown for fast rotation (Earth-like, in the upper row) and the much slower pseudo-synchronous rotation, defined in Equation \ref{eq:PSR}, in the lower), at orbital eccentricities of 0.3 and 0.6. The dark dotted lines delineate the star-facing hemisphere, which is centered in each plot.}
\label{fig:CLOUD_globes}
\end{center}
\end{figure*}

Given these strong day-night contrasts in humidity, it is not surprising that the slow rotators at both eccentricities show extremely large cloud water paths (vertically integrated condensed water in cloud droplets), approaching 1 kg m$^{-2}$ following periastron\footnote{For comparison, the same mass surface density in Earth's tropics peak near 200 g m$^{-2}$ \citep{ODell2008}.}, centered on the sub-stellar point at periastron (Figure~\ref{fig:CLOUD_globes}), and largely mimicking the features seen in both surface temperature and albedo.  The negative cloud forcing in the shortwave is presumably responsible for limiting sub-stellar surface temperatures at periastron, causing the peak ocean temperatures to exist in a partial ring around the sub-stellar point (Figure~\ref{fig:ALBEDO_globes}).

Taken together, all of these quantities show that slowly rotating planets on orbits of modest to high eccentricity can become mostly ice covered, except for a longitudinally confined warm, cloudy, and ice-free region region that persists at low latitudes through the orbit.  This is consistent with previous simulations of synchronously rotating aquaplanets on circular orbits \citep[e.g.][]{jos97, mer10} and on eccentric orbits \citep{bol16}, except that the ice-free region does not remain on the day side at apastron for pseudo-synchronous rotation (see \S \ref{sec:results:external:observables} for further discussion).  
The implications of this shift in the warm, ice-free region for observable emission is further complicated by the fact that atmospheric temperatures vary much more than surface temperatures over the orbit (e.g.\ Figure~\ref{fig:Q-P_T-P}), responding to the orbital cycle of instellation.  In contrast, for faster, Earth-like rotation, all quantities are homogenized longitudinally, so that outgoing radiation will be set by the orbital variations in stellar heating.

\subsection{External Results}\label{sec:results:external}
\subsubsection{Outgoing Longwave Radiation}\label{sec:results:external:OLR}
\begin{figure*}
\begin{center}
   \begin{tabular}{cc}
     \includegraphics[width=8.5cm]{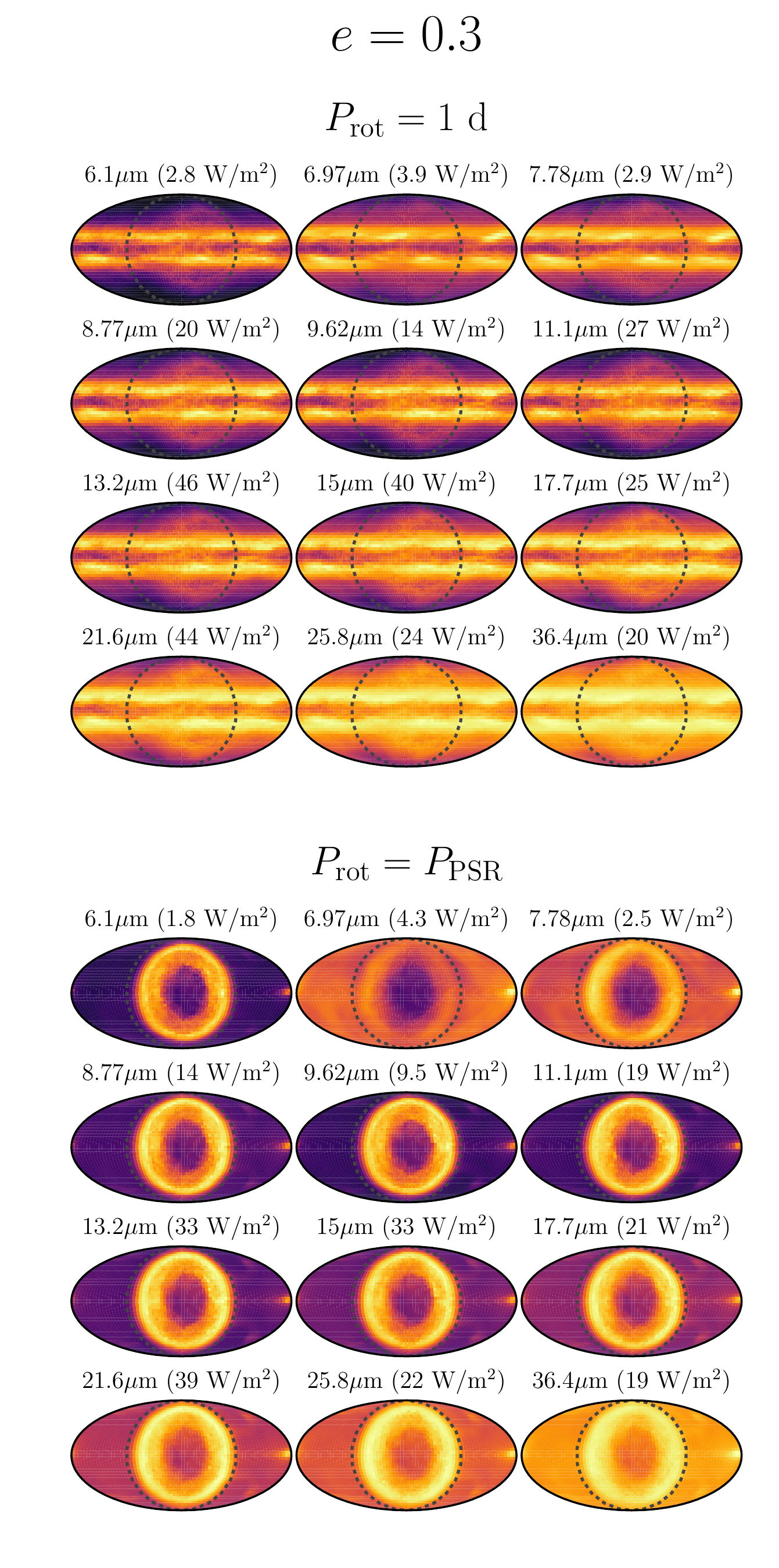} &
     \includegraphics[width=8.5cm]{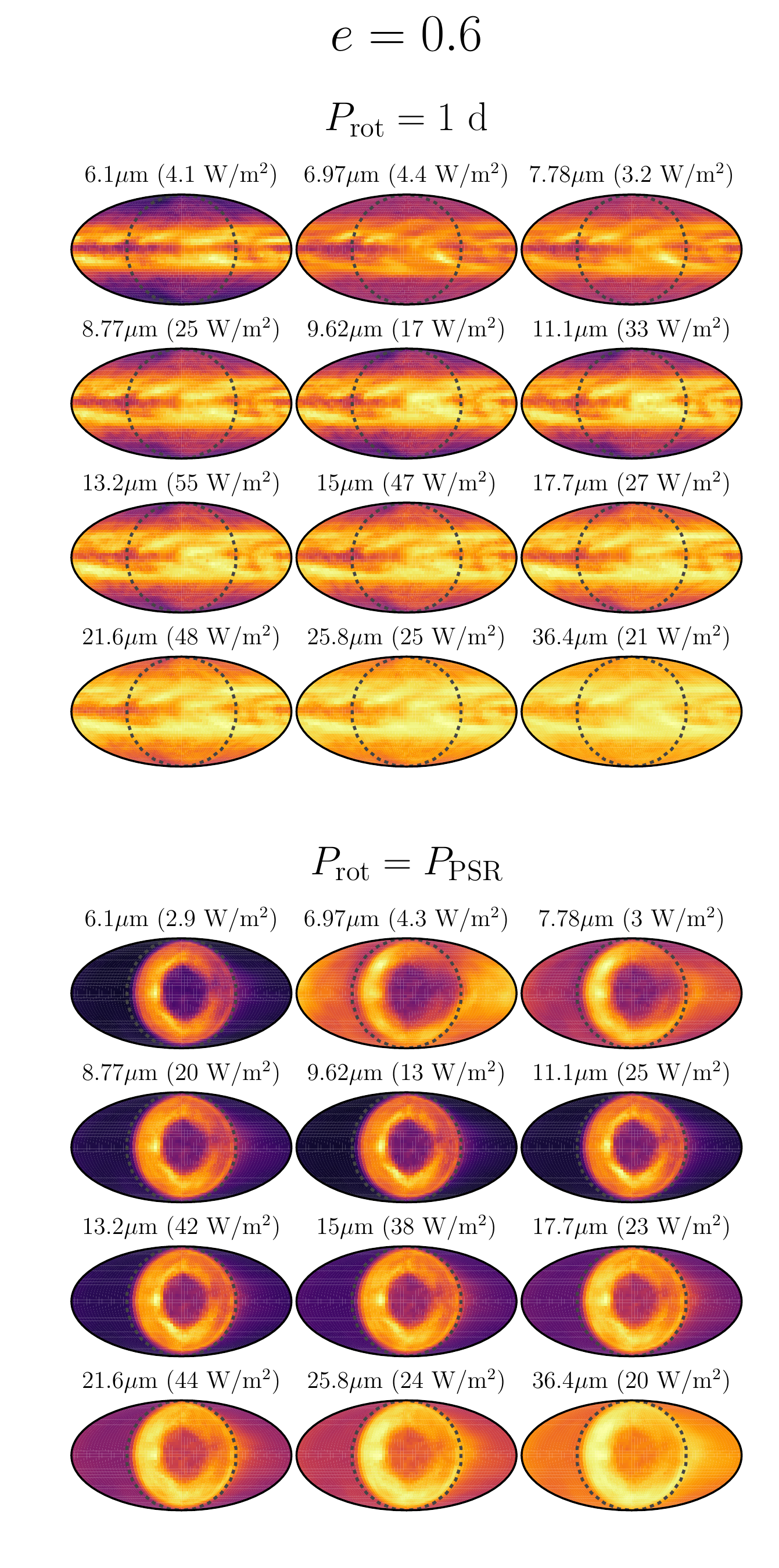} \\
   \end{tabular}
\caption{Global outgoing flux at the top of the atmosphere for a range of infrared bands from the ExoCAM model (Bands 7--18 in Table \ref{table:spectral_bands}), for each of the two rotation periods and orbital eccentricities, during periastron. The dotted lines delineate the star-facing hemisphere, which is centered in each plot. Each sub-plot has a color range with zero flux as black, and the brightest color given by the flux above each globe in parentheses.}
\label{fig:OLR_spectrum_globes}
\end{center}
\end{figure*}

As an intermediate step between characterizing the atmospheric state and simulating what might be observed by a telescope, we examine the horizontal distribution of planetary radiative emission in the spectral bins used by the model radiation scheme.  The peak radiative flux in each spectral bin is similar (within $\pm$50\%) across rotation rates and orbital phase.  The fast rotators have emission that is nearly uniform in longitude with large equator-to-pole gradients, as expected.  In contrast, the night-side fluxes of the slow rotators at most wavelengths remain much lower when compared to either the slow-rotation day side fluxes or the fluxes in the equatorial regions of the fast rotators (Figure~\ref{fig:OLR_spectrum_globes}).  In all bands, the slow-rotation sub-stellar point is an emission minimum due to the thick cloud shield, and the maximum emission occurs in a ring closer to the edge of the day side.

The contrast between night side emission and the emission from this ring at the edges of the day side is weaker in the 6.97 and 7.78 $\mu$m bands, which both lie in the water vapor absorption band that is centered at 6.3 $\mu$m (and which, in turn, absorbs strongly between 5 and 8 $\mu$m).  In these spectral regions, the dryness of the atmosphere on the cold night side allows radiation to escape from the lower troposphere or the surface itself, while the high optical depth on the warm day side allows emission only from higher (and thus colder) levels of the upper troposphere. The day-night contrast for slow rotation also weakens for the same reasons as we move out beyond 20 $\mu$m, into the reddest bands occupied by the rotational absorption features of water.

These emission distributions show that the strong longitudinal temperature contrasts in our slow rotators are best observed in longwave bands away from the major water vapor absorption features.  When considering observables, it must also be borne in mind that for a non-tidally locked planet, the warm side will not always be the day side (e.g.\ Figure~\ref{fig:ALBEDO_globes}).  Emission changes caused by planet-star distance variations over an orbit may also complicate the inferences that can be made about rotation rate from observables.  This motivates construction of simulated light curves in the next section.

\subsubsection{Observable Light Curves}\label{sec:results:external:observables}
\begin{figure*}
\begin{center}
   \begin{tabular}{c}
     \includegraphics[width=15cm]{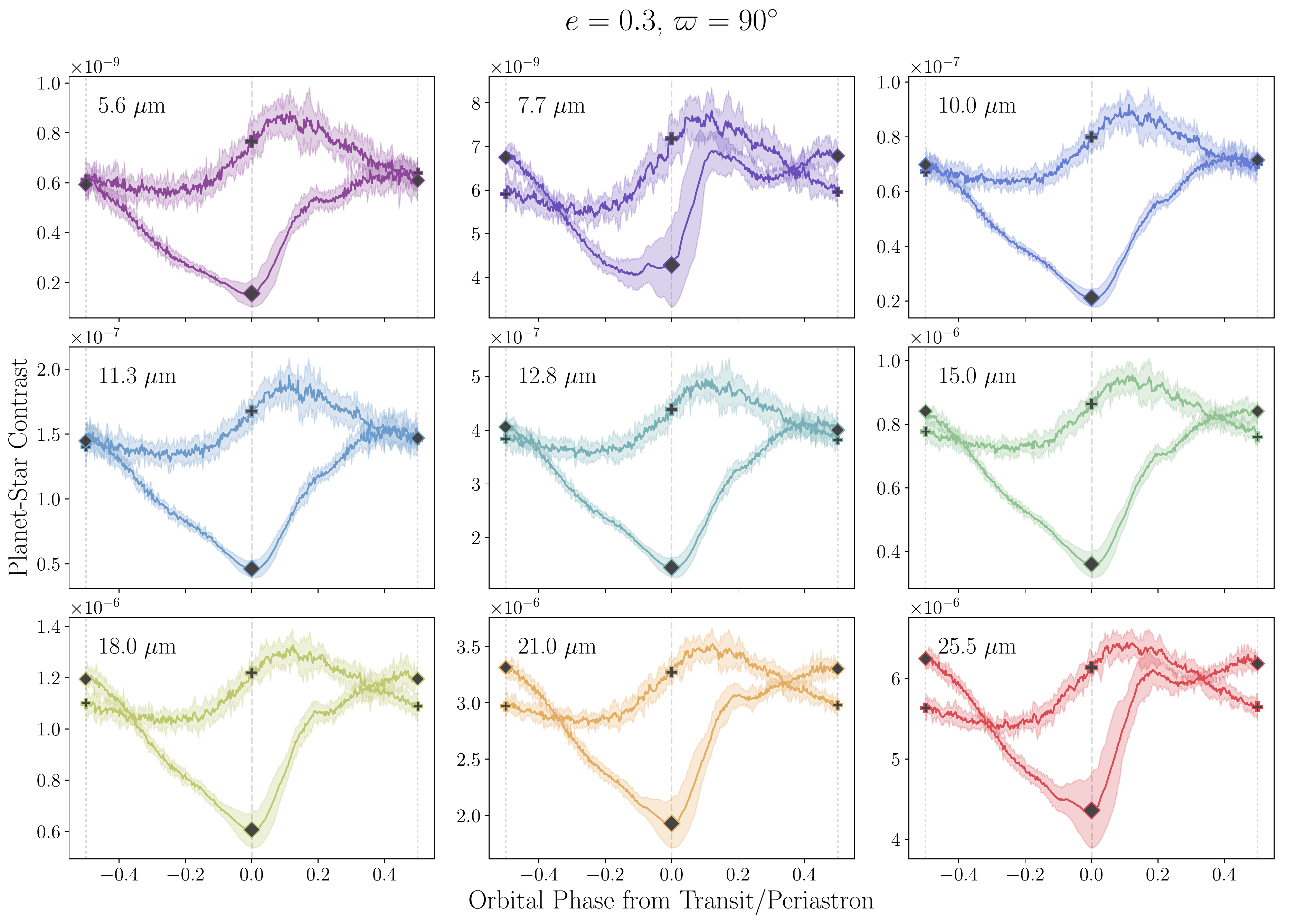} \\
     \includegraphics[width=15cm]{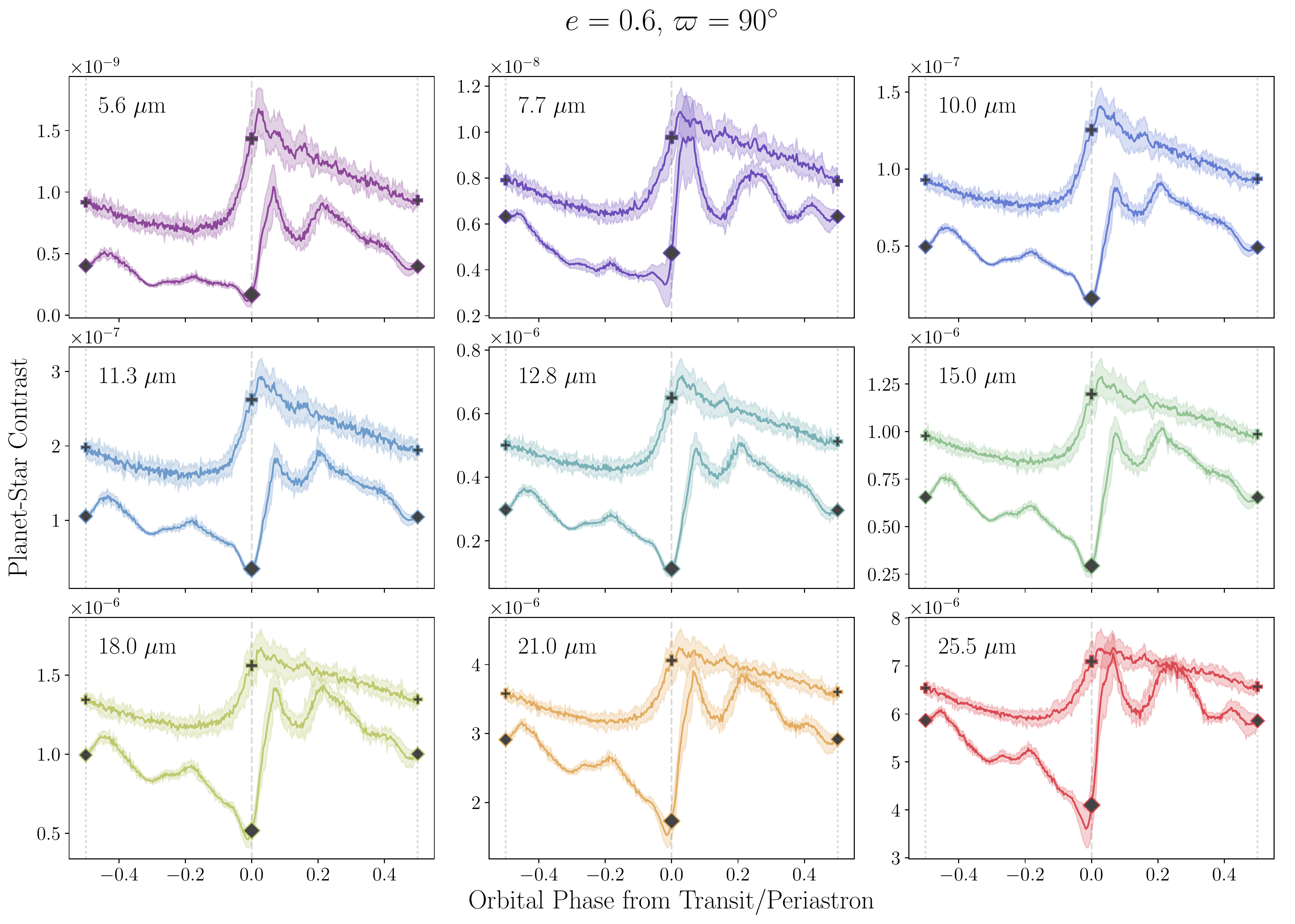} \\
   \end{tabular}
\caption{Simulated light curves for a viewing geometry such that the orbit is seen edge-on, and the planet transits during periastron. Each 3x3 grid represents one of the two orbital eccentricities. The light curves are plotted for each of the 9 MIRI bands of the upcoming James Webb Space Telescope. Within each plot, the light curve with plus-sign markers shows Earth-like (fast) rotation, and the curve with diamond markers shows pseudo-synchronous (slow) rotation. The solid color lines are the averages over the final 10 orbits, and the surrounding shaded region represents the range of fluxes over the orbits.}
\label{fig:MIRI_lightcurves_omega90}
\end{center}
\end{figure*}

\begin{figure*}
\begin{center}
   \begin{tabular}{c}
     \includegraphics[width=15cm]{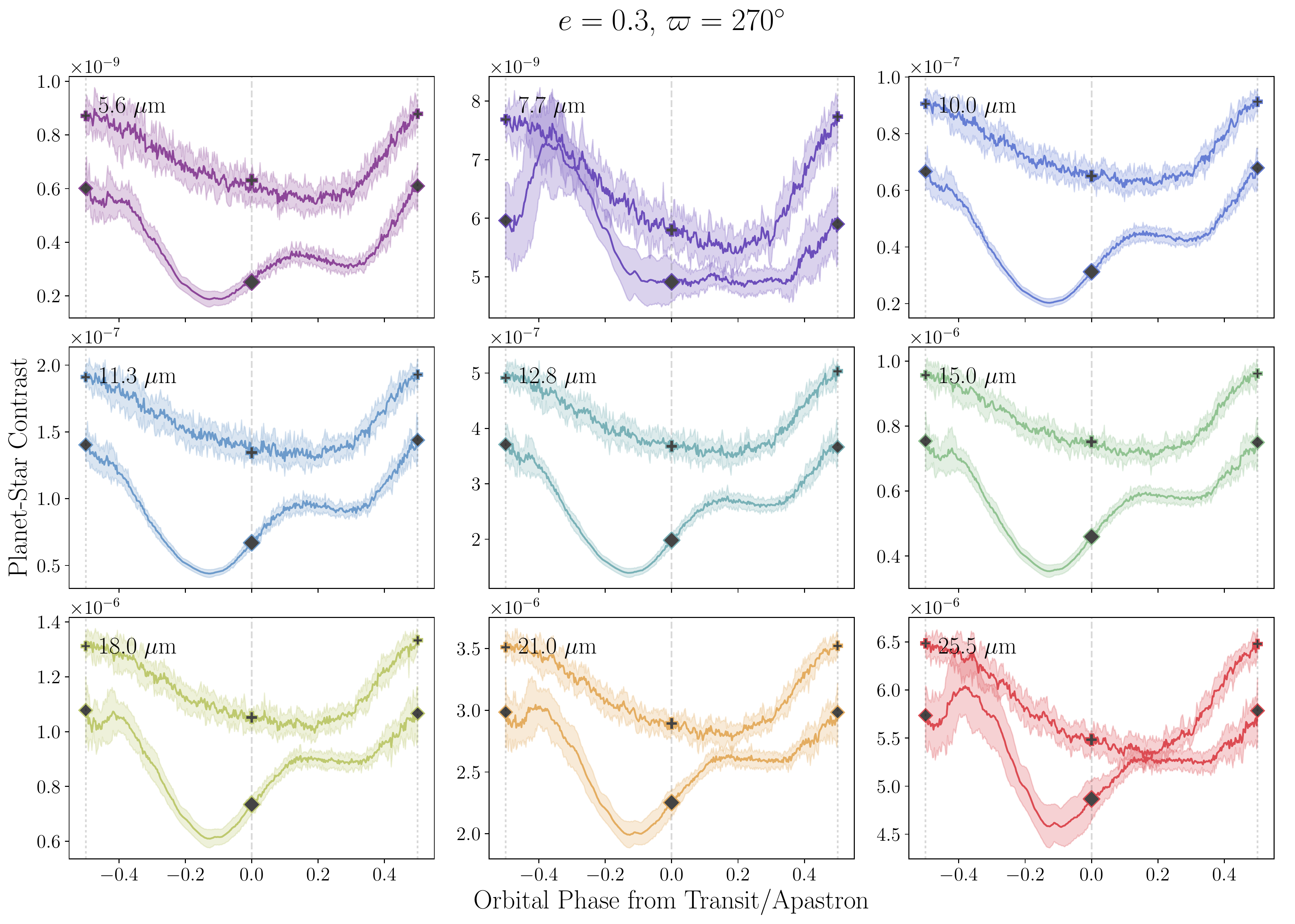} \\
     \includegraphics[width=15cm]{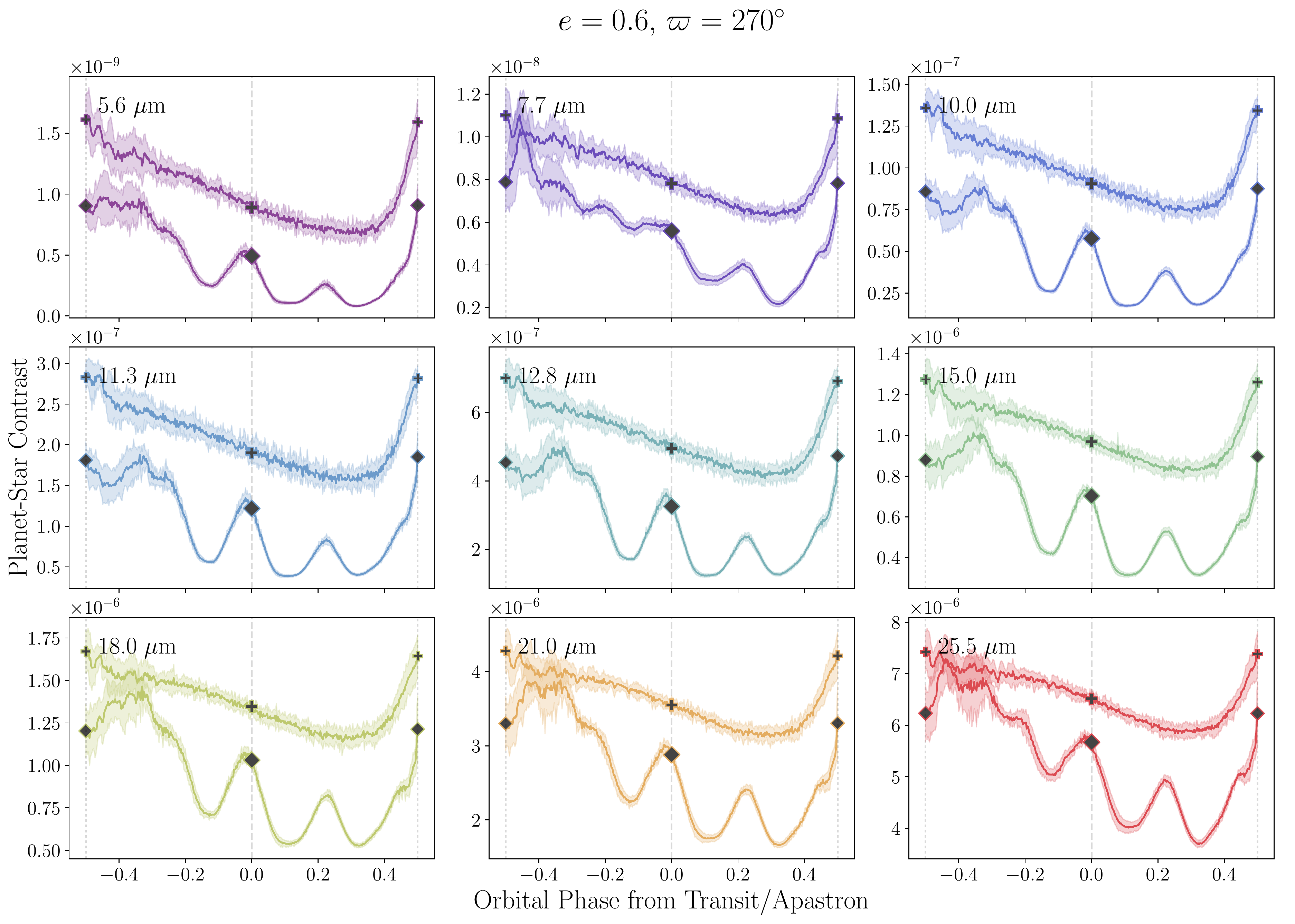} \\
   \end{tabular}
\caption{Simulated light curves for a viewing geometry such that the orbit is seen edge-on, and the planet undergoes secondary eclipse during periastron. Each 3x3 grid represents one of the two orbital eccentricities. The light curves are plotted for each of the 9 MIRI bands of the upcoming James Webb Space Telescope. Within each plot, the light curve with plus-sign markers shows Earth-like (fast) rotation, and the curve with diamond markers shows pseudo-synchronous (slow) rotation. The solid black lines are the averages over the final 10 orbits, and the surrounding shaded region represents the range of fluxes over the orbits.}
\label{fig:MIRI_lightcurves_omega270}
\end{center}
\end{figure*}

Adopting the methodology of converting model output to photometry, we generate light curves based on the JWST MIRI filter profiles, for the two assumed viewing geometries described in \S \ref{sec:model:observables}. Figure~\ref{fig:MIRI_lightcurves_omega90} shows the model light curves at both eccentricities for a viewing geometry where the planet's transit and periastron passage coincide ($\varpi=90^\circ$), and Figure~\ref{fig:MIRI_lightcurves_omega270} shows light curves for a geometry where eclipse and periastron passage coincide ($\varpi=270^\circ$). We discuss four primary qualities of these light curves:
\begin{enumerate}
\item For Earth-like rotation, the variation follows the expectation for a longitudinally symmetric system in which the flux follows the variations in planet-star distance over the orbital cycle, modified by the viewing geometry. Accordingly, the light curve variations have significantly higher amplitude at higher eccentricity (note the change in vertical scale between eccentricities in Figures~\ref{fig:MIRI_lightcurves_omega90} and \ref{fig:MIRI_lightcurves_omega270}). The light curves for fast rotators reach their peak at or shortly after apastron, when the planet is closest to the star and thus hottest, and the light curve morphology is largely independent of wavelength.

\item For pseudo-synchronous rotation, the phase variations are more complicated but are generally affected strongly by the longitudinal temperature contrasts. Where eclipse aligns with periastron and thus the observer sees the day side (corresponding to $\varpi=270^\circ$, in Figure~\ref{fig:MIRI_lightcurves_omega270}), the light curves have maxima at or just after periastron, giving the slow-rotation light curves similar times of extrema to the fast rotators for both eccentricities. Conversely, where eclipse is aligned with apastron ($\varpi=90^\circ$, Figure~\ref{fig:MIRI_lightcurves_omega90}) the light curves reach an absolute minimum near periastron.  For $e$ = 0.3 the slow-rotation light curves are nearly 180\dgs out of phase with the light curves for Earth-like rotation.  In contrast, for $e$ = 0.6 the flux quickly brightens after the periastron minimum as the rotation in this part of the orbit brings the highly irradiated hemisphere into view, resulting in the maxima of the slow-rotation light curves occurring at nearly the same orbital phase as the maxima of light curves for fast rotation.

\item The fluxes of the slow rotator are consistently lower than their fast counterparts for $\varpi=270^\circ$ and for the high-eccentricity curves with $\varpi=90^\circ$. This is consistent with the slow-rotation cases having ice-covered sides substantially colder than the mean temperature of the fast rotators (Figure~\ref{fig:ALBEDO_globes} and Table \ref{table:surface_temperatures}), as well as much higher upper-tropospheric humidities (Figure~\ref{fig:Q-P_T-P}) on the warm sides of the planets, and therefore much higher cloud water paths (Figure~\ref{fig:CLOUD_globes}).

\item The simple thermal expectation is that, as the bands move toward longer wavelengths, the contrast ratio will increase. We see this in both rotation rates, but additional wavelength-dependent features are present for slow rotation in particular that affect the shapes of the periastron-induced maxima for $\varpi=270^\circ$. For example, in the $e=0.3$ light curves for the slow rotator, there are peaks shortly after periastron consistent with H$_2$O emission in the F770W band (centered at 7.7 $\mu$m) and at wavelengths longer than 18 $\mu$m.  The order of magnitude increase in upper-atmosphere moisture content (as seen in Figure~\ref{fig:Q-P_T-P}) provides the slower rotators with the water vapor needed to intensify the observed flux in this band.
\end{enumerate}

Beyond these major features, some additional qualities of the light curves warrant explanation. Secondary peaks occur in most of the slow-rotation light curves and come from the ratio of the rotation period to the orbital period. This creates a single secondary peak for $e=0.3$, where the ratio is $\approx 0.64$ (i.e.\ roughly 2 rotations each orbital period), and 3 secondary peaks for $e=0.6$, where the rotation period is close to one-quarter the orbital period. This effect is often referred to as ``ringing'', and is due to the day-side hemisphere from periastron passage retaining a high temperature as it rotates in and out of view. Such an effect was predicted in the hydrodynamical simulations of \citet{Langton2008} for planets such as HD 80606 b, and further seen in the models of \citet{cow11a} and \citet{Kataria2013}. The ringing only comes about for strong enough day-night temperature differences, and if both the thermal timescale and rotation period are shorter (though not significantly shorter) than the orbital period; this condition was used to constrain the rotation period of the highly eccentric HD 80606 b, whose phase variations have been observed not to exhibit this ringing behavior \citep{dew16,lew17}. 

Finally, we note that the inter-orbital variability in flux is generally broader and more consistent over orbital phase for the fast rotators. For the slow rotators, the highest inter-orbital variability is found around periastron, and overall is stronger for the water-sensitive bands. While these differences in inter-orbit variability could provide a potential probe for distinguishing rotation rate, a more in-depth analysis of this effect would require simulating a much larger number of orbits than what we have presented here. We leave such an analysis for future work.

\subsubsection{Constructing Light Curve Ratios}\label{sec:results:external:LightCurveRatios}
\begin{figure*}
\begin{center}
   \begin{tabular}{cc}
     \includegraphics[width=8.5cm]{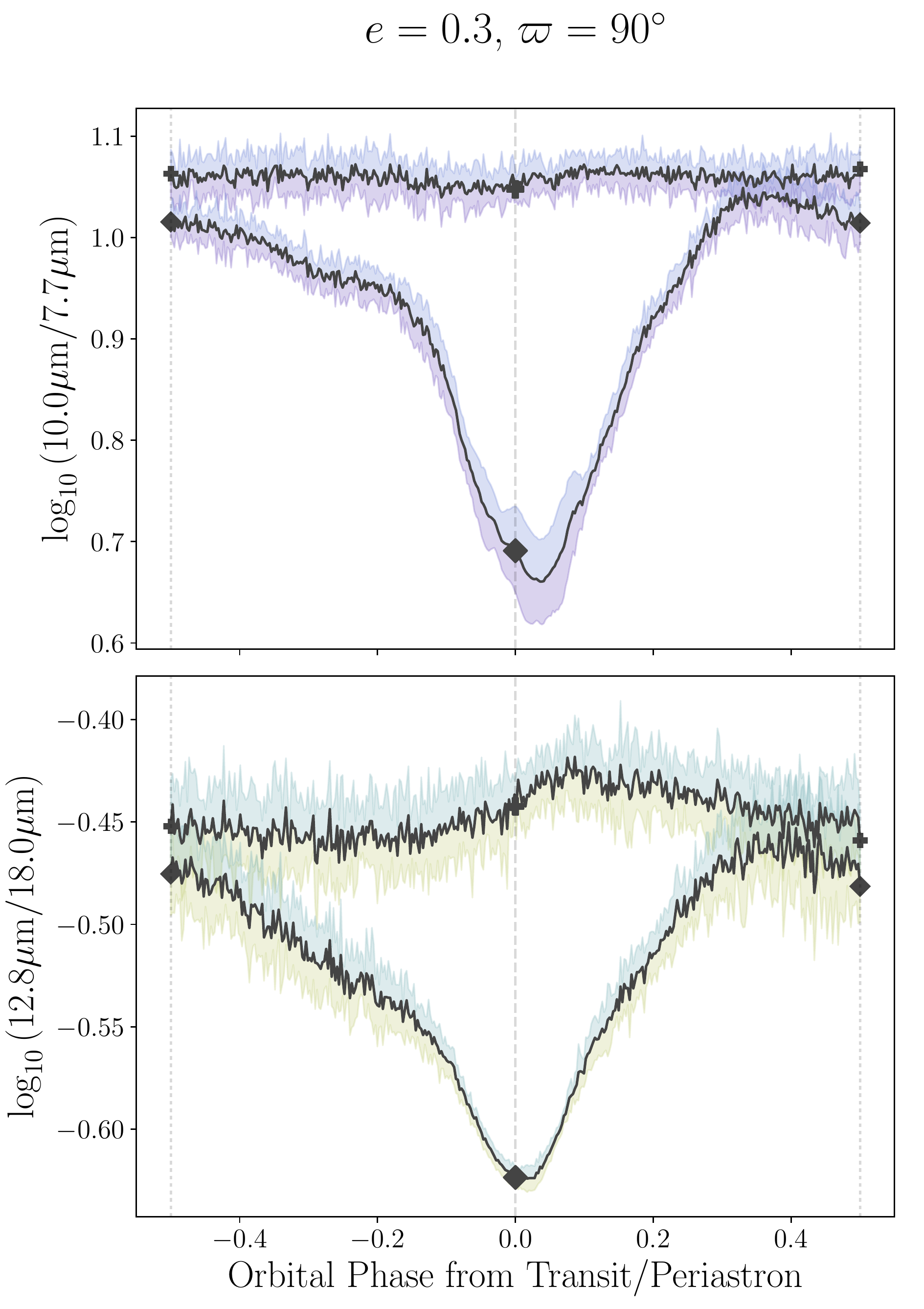} &
     \includegraphics[width=8.5cm]{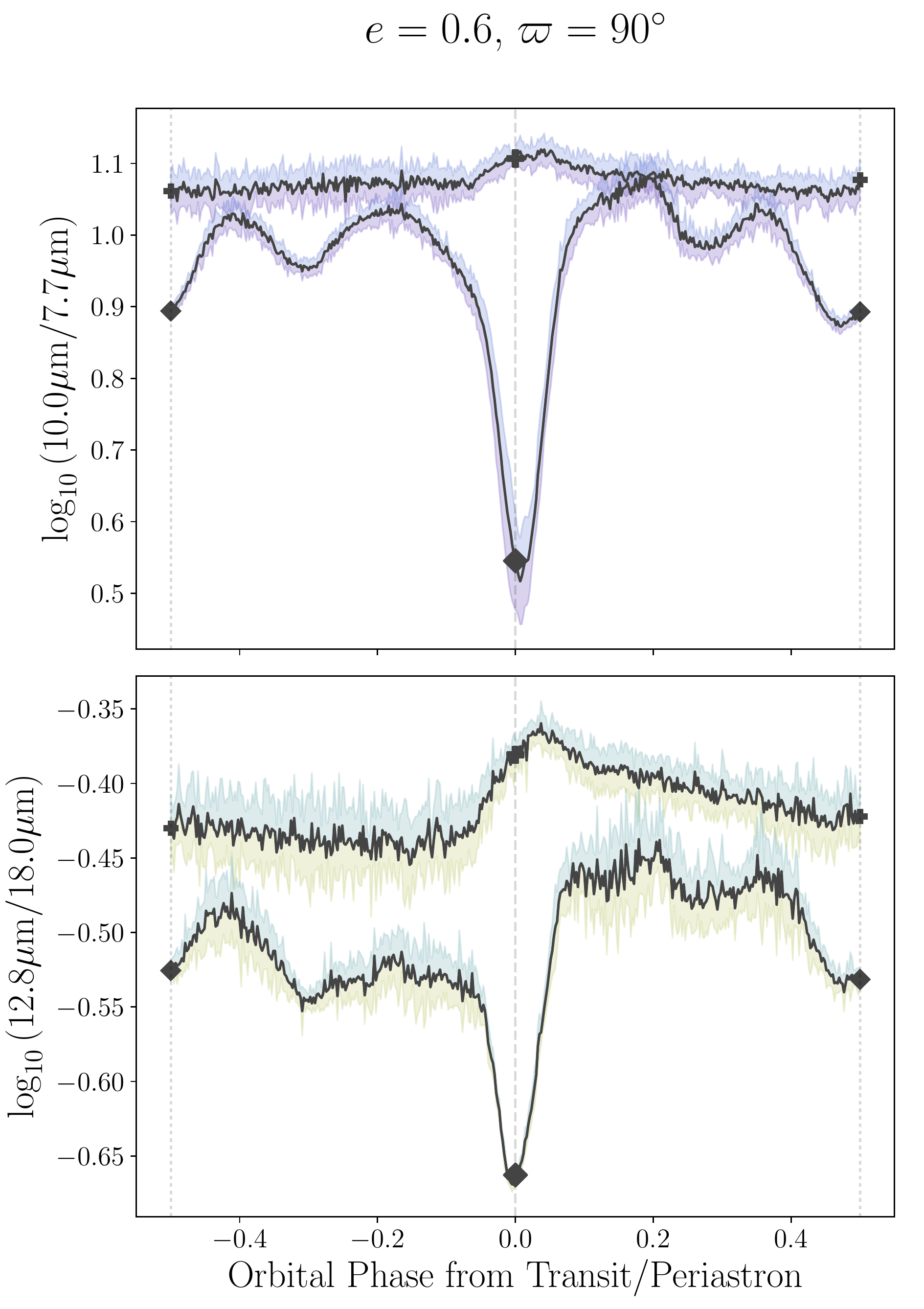} \\
   \end{tabular}
\caption{Ratios of a selection of the planet-star contrasts shown in the light curves of Figure~\ref{fig:MIRI_lightcurves_omega90}, where the planet transits during periastron. Within each plot, the light curve with plus-sign markers shows Earth-like (fast) rotation, and the curve with diamond markers shows pseudo-synchronous (slow) rotation. The solid black lines are the averages over the final 10 orbits, and the surrounding shaded region represents the range of fluxes over the orbits. The choice of colors for these regions is purely to illustrate the fluxes used for each ratio.}
\label{fig:MIRI_ratio_omega90}
\end{center}
\end{figure*}

\begin{figure*}
\begin{center}
   \begin{tabular}{cc}
     \includegraphics[width=8.5cm]{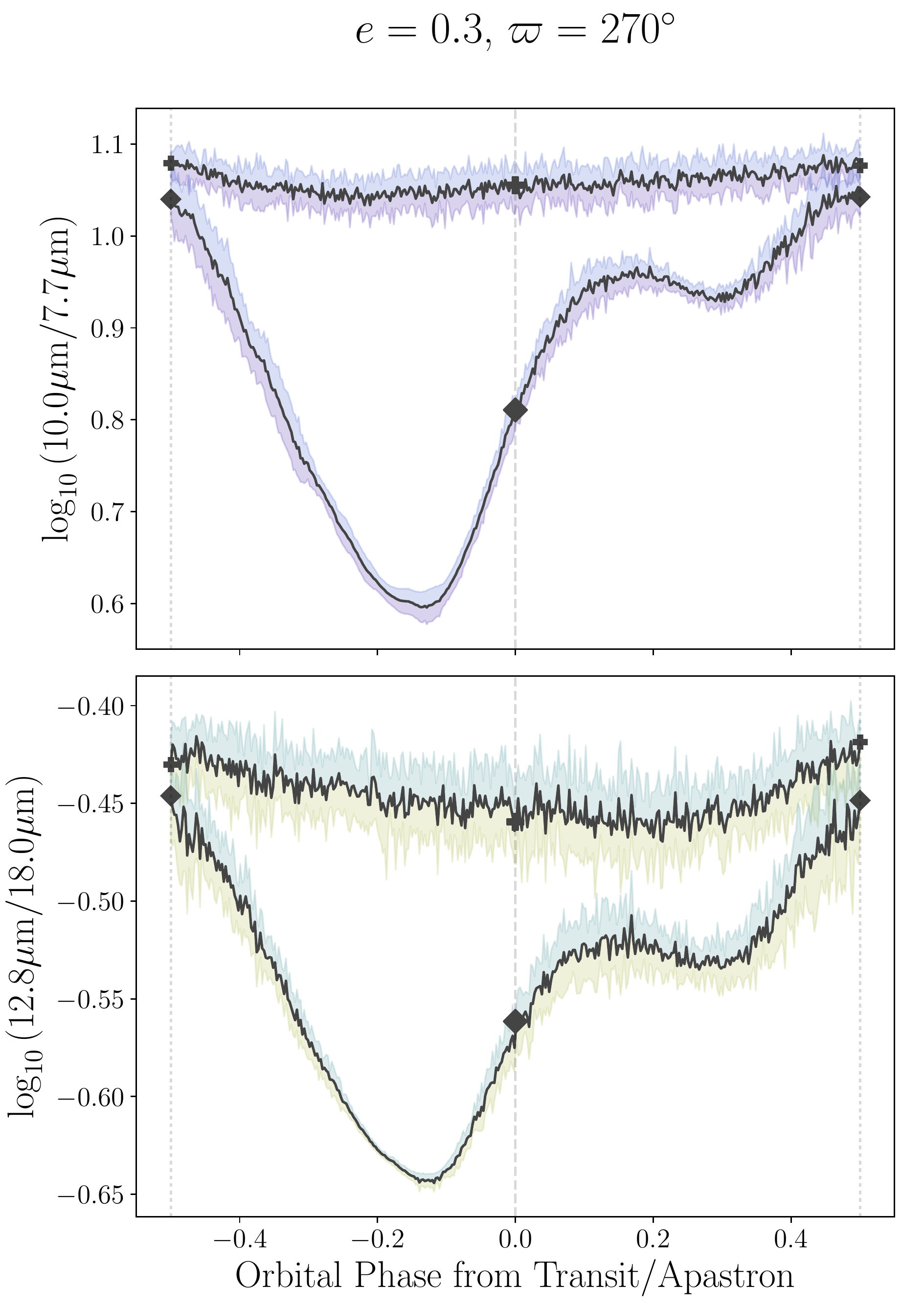} &
     \includegraphics[width=8.5cm]{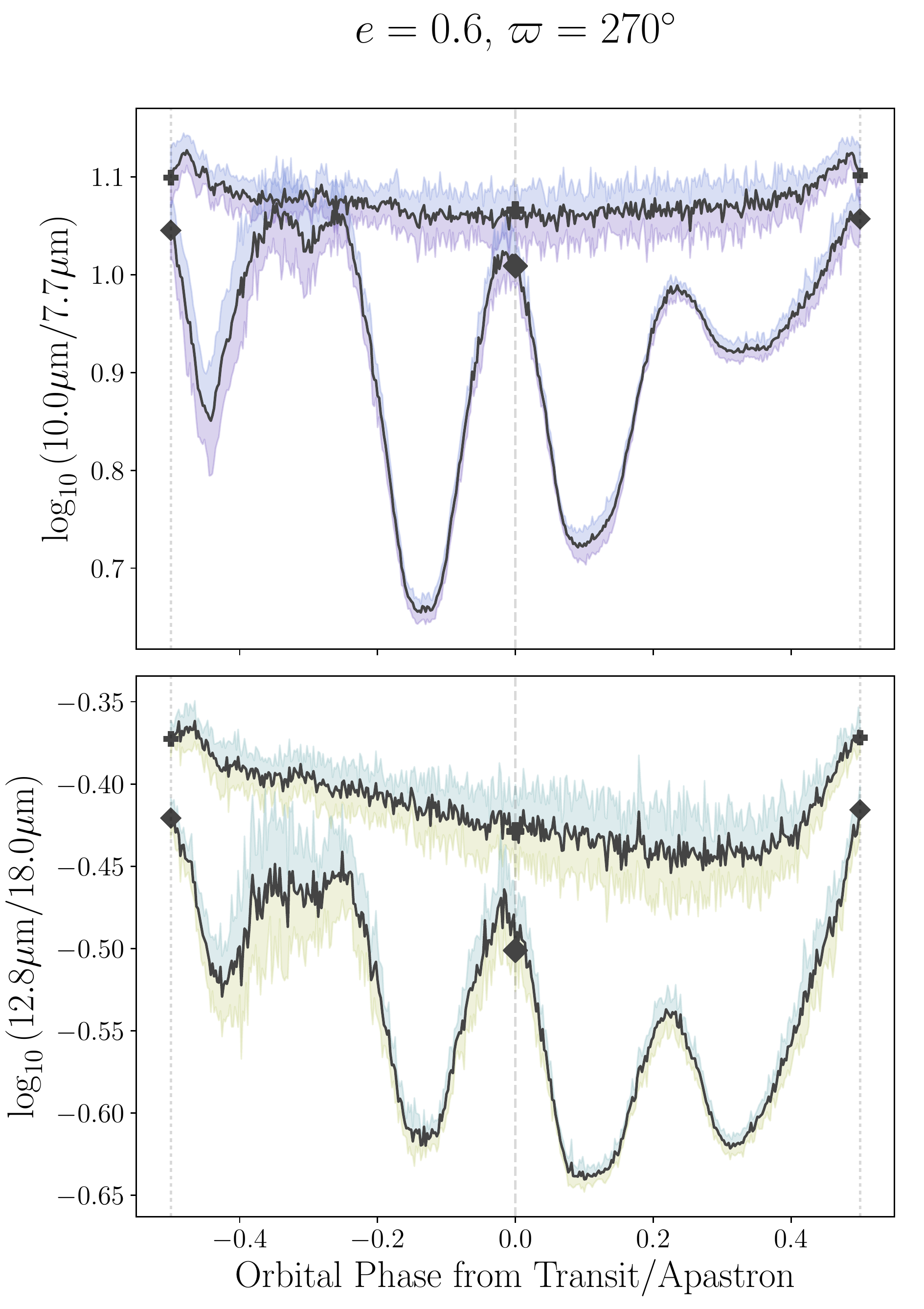} \\
   \end{tabular}
\caption{Ratios of a selection of the planet-star contrasts shown in the light curves of Figure~\ref{fig:MIRI_lightcurves_omega270}, where the planet transits during apastron. Within each plot, the light curve with plus-sign markers shows Earth-like (fast) rotation, and the curve with diamond markers shows pseudo-synchronous (slow) rotation. The solid color lines are the averages over the final 10 orbits, and the surrounding shaded region represents the range of fluxes over the orbits. The choice of colors for these regions is purely to illustrate the fluxes used for each ratio.}
\label{fig:MIRI_ratio_omega270}
\end{center}
\end{figure*}

With all of the above features in mind, we construct ``colors'' by comparing the fluxes in two bands. Here we choose two pairs of bands: the bands at 7.7 and 10.0 $\mu$m, and the bands at 12.8 and 18.0 $\mu$m (Figures \ref{fig:MIRI_ratio_omega90}--\ref{fig:MIRI_ratio_omega270}).  Each of these pairs compares one band where water has a strong absorption feature with another band in the water vapor window.  For the first pair, which consist of shorter wavelengths, the longer-wavelength band is in the water vapor window while the shorter lies in the 6.3 $\mu$m water vapor vibrational-rotational absorption band.  The converse is true of the second pair of bands, which lie at longer wavelengths:  the shorter-wavelength band is in the water vapor window and the longer band lies in the short-wavelength end of the pure rotational, far-infrared absorption band of water vapor.  The behavior of these colors are discussed in greater detail in Appendix \ref{sec:appendix:colors}.

For a transit-periastron viewing angle ($\varpi=90^\circ$) we see a consistent difference in the colors during or near periastron. The differences can be as high as 0.35 dex for both eccentricities. At $\varpi=270^\circ$, the peak difference between slow and fast rotators is not well-aligned with periastron for $e=0.3$, reaching a maximum notably prior to transit and having a smaller secondary maximum mid-way between transit and eclipse. At $e=0.6$ and $\varpi=270^\circ$, the pseudo-synchronous rotation period is short enough to allow a spike in each color near periastron for the slow rotator; large differences with the fast rotator thus occur about four times throughout the orbit consistent with the roughly 4:1 ratio of rotational to orbital periods. The $e=0.6$ color curves exhibit some variations from orbit to orbit, but one persistent feature is the set of secondary dips/peaks which correspond to the spin-orbit ratio.

\begin{figure*}
\begin{center}
   \begin{tabular}{cc}
     \includegraphics[width=8.5cm]{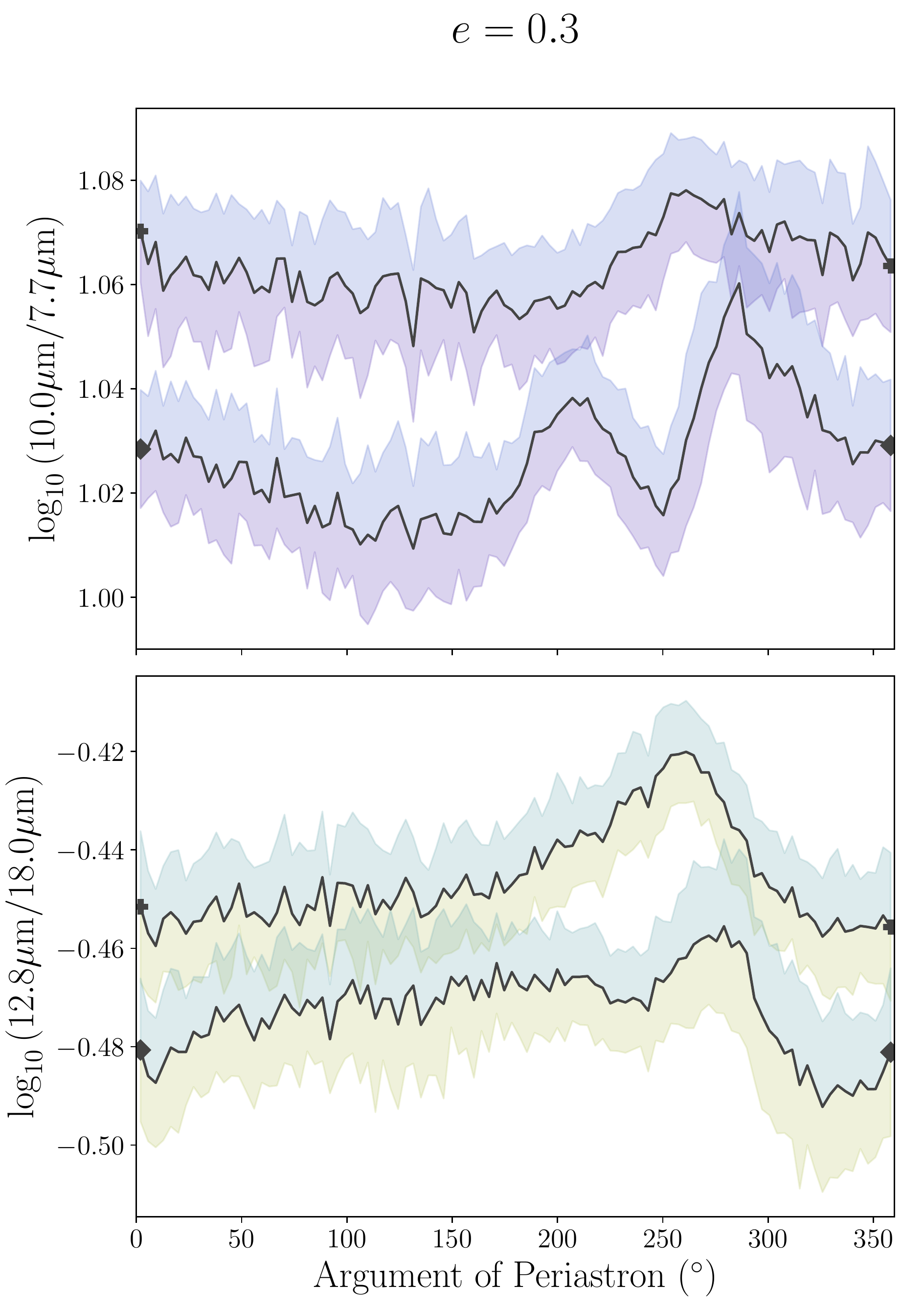} &
     \includegraphics[width=8.5cm]{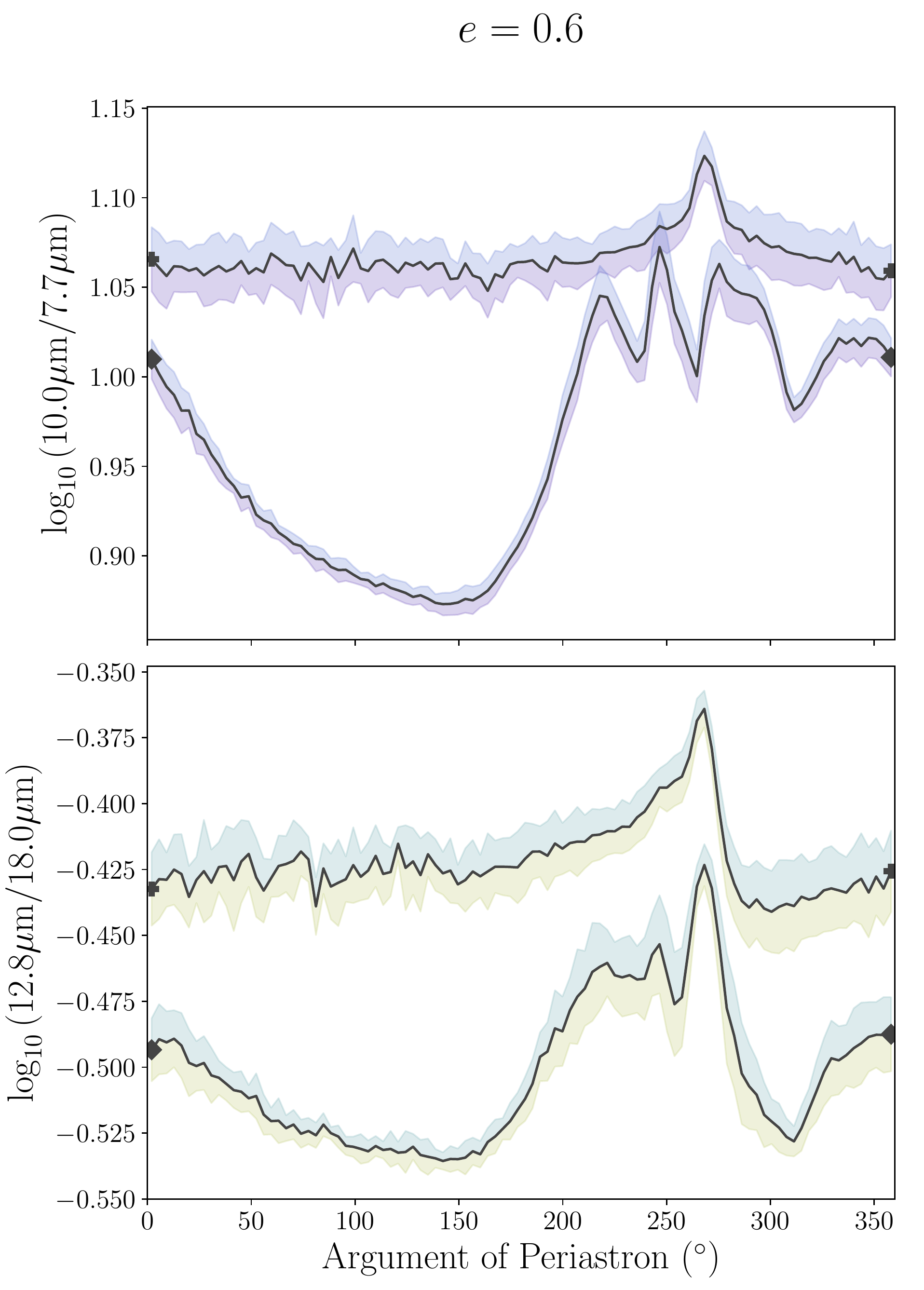} \\
   \end{tabular}
\caption{Predicted eclipse depth ratios for the range of possible longitudes of periastron relative to an observer. The quantities plotted are identical to those in Figures \ref{fig:MIRI_ratio_omega90} and \ref{fig:MIRI_ratio_omega270}. The solid color lines are the averages over the final 10 orbits, and the surrounding shaded region represents the range of fluxes over the orbits. The choice of colors for these regions is purely to illustrate the fluxes used for each ratio.}
\label{fig:MIRI_ratio_eclipsedepths}
\end{center}
\end{figure*}

\begin{figure*}
\begin{center}
   \begin{tabular}{cc}
     \includegraphics[width=8.5cm]{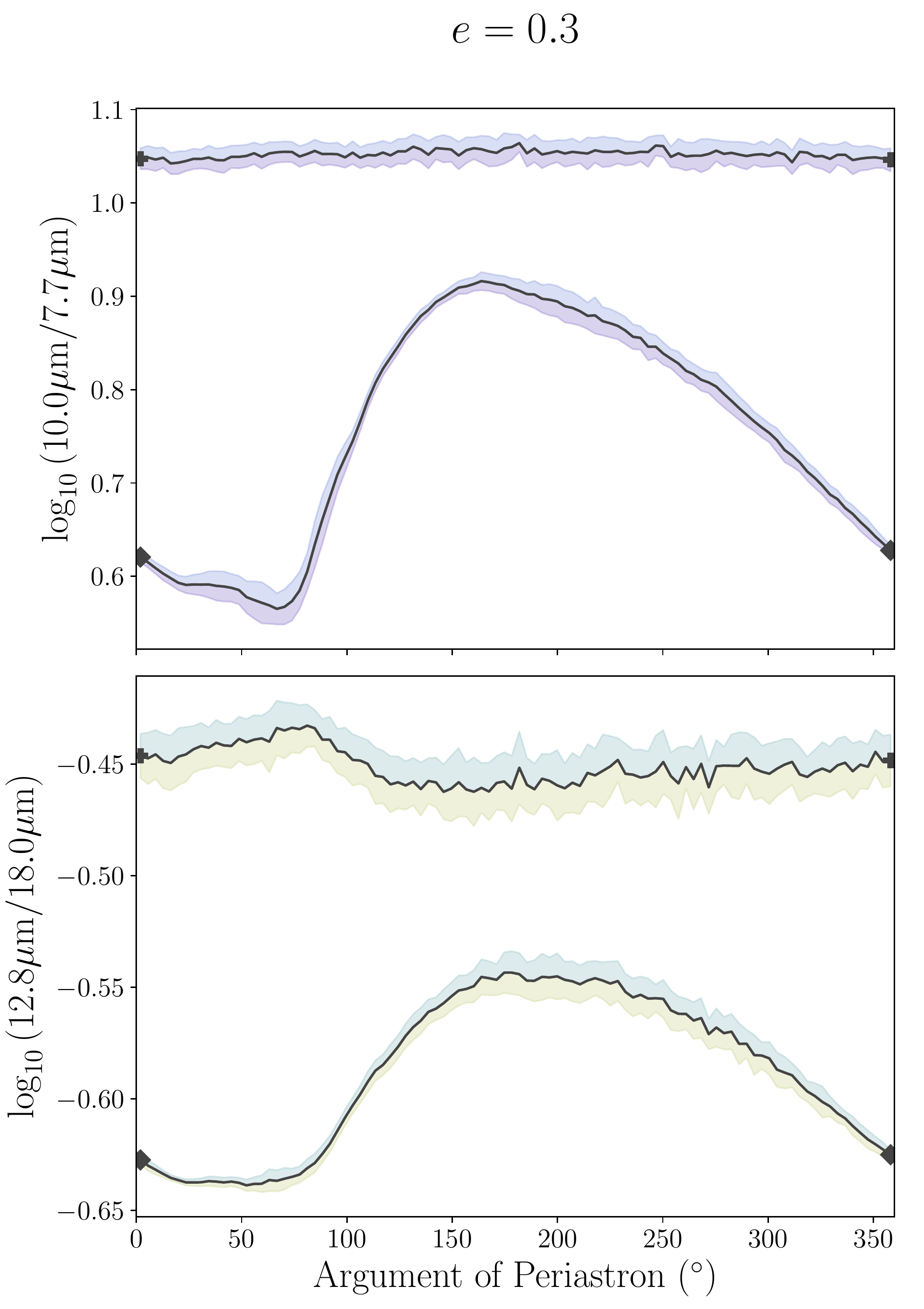} &
     \includegraphics[width=8.5cm]{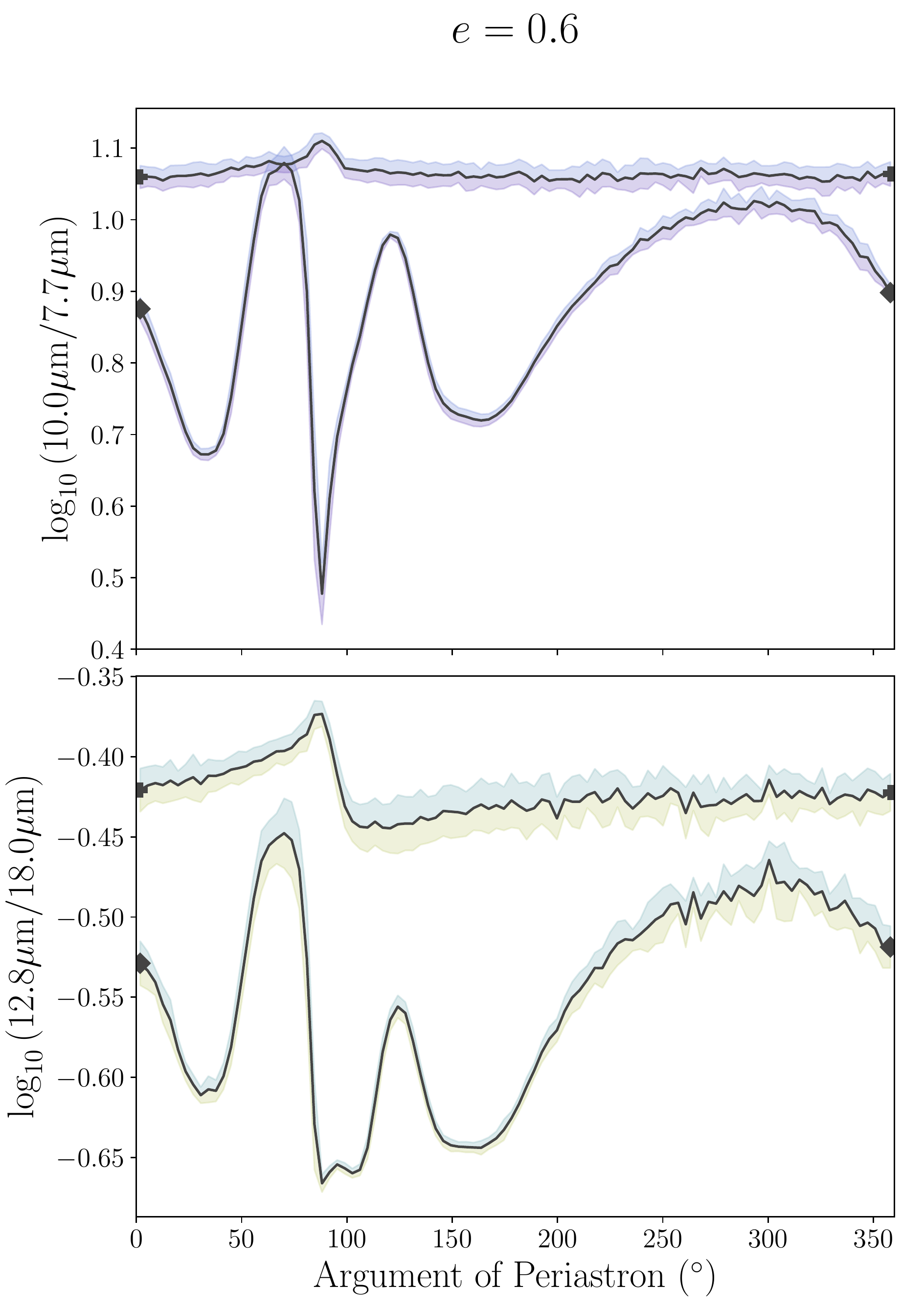} \\
   \end{tabular}
\caption{Predicted ratios of the observed night-side contrasts during transit for the range of possible longitudes of periastron relative to an observer. The quantities plotted are identical to those in Figures \ref{fig:MIRI_ratio_omega90} and \ref{fig:MIRI_ratio_omega270}. The solid color lines are the averages over the final 10 orbits, and the surrounding shaded region represents the range of fluxes over the orbits. The choice of colors for these regions is purely to illustrate the fluxes used for each ratio.}
\label{fig:MIRI_ratio_transitdepths}
\end{center}
\end{figure*}

Despite limiting our analysis of full-orbit photometry to two extreme cases of $\varpi = 90^\circ$ and $270^\circ$, it is relatively straightforward to predict the eclipse depths for the entire range of possible observing longitudes (Figure~\ref{fig:MIRI_ratio_eclipsedepths}). The variations in eclipse depths in both cases with longitude only show minor variations relative to the ranges seen in Figures \ref{fig:MIRI_ratio_omega90} and \ref{fig:MIRI_ratio_omega270}. Given this, it would be comparatively difficult to distinguish the scale of rotation from eclipse depths alone; therefore we also examine the night-side fluxes one would observe during transit (Figure~\ref{fig:MIRI_ratio_transitdepths}). Here we gain the advantage of the strong day-night water-induced contrasts seen in the slow rotation cases. From these, we suggest that observations of day-side fluxes during/near eclipse, coupled with night-side fluxes during/near transit, could help discern these two cases.

The variations in eclipse depths with viewing geometry show a similar qualitative behavior as the phase curves with respect to rotation: the fast rotators exhibit a much weaker dependence on the observing angle than the slow rotators. This further suggests that, while maximizing the observing time would maximize the ability to discern between these cases, for a wide range of viewing geometries a pair of eclipse depths could hint at the broad timescale of rotation.

\section{Discussion}\label{sec:discussion}
We showed that, for simulated ocean-covered planets on eccentric orbits, differences in the scale of rotation rate could be discernible from mid-infrared phase photometry of sufficient signal-to-noise to detect a contrast of at least 1 ppm relative to the host star. We explicitly modeled two contrasting viewing geometries and demonstrate that differences of order $\sim\!0.3$--0.4 dex in photometric contrast ratios would be distinguishable with a combination of broadband flux observations during transit and secondary eclipse. These differences are caused by the strong dependence of phase curve features, at particular wavelengths, on the concentration of upper-tropospheric water vapor. We have shown that, while any day-night contrasts in the temperature are restricted to layers near the planets' surfaces, day-night contrasts in atmospheric moisture concentration become very strong well into the upper atmosphere for rotation with periods comparable with the orbital period. These characteristic contrasts impart significant additional variations in those specific photometric bands which contain water absorption features, most notably in the MIRI band centered at 7.7 $\mu$m, but also broadly in the mid-infrared beyond approximately 20 $\mu$m.

Our results have used the instrumental responses for the upcoming James Webb Space Telescope to predict needed future mid-infrared sensitivity. The photometric precision required to discern the predicted variations are quite small, beyond what JWST will be capable of even with the most optimistic expectations. Looking forward to proposed space-based missions in this wavelength range, the Origins Space Telescope (OST) \citep{Battersby2018} will have coverage in the mid-infrared with its Mid-Infrared Imager, Spectrometer, Coronagraph (MISC) instrument suite \citep{Sakon2018}. With an estimated precision of $\sim\!5$ ppm in the range $\sim\!3$--20 $\mu$m, OST is designed to have the potential to observe thermal phase variations of terrestrial planets \citep{Fortney2018,Kataria2018}. Therefore, our findings point toward achieving these sorts of observations with a future generation of space observatories, perhaps first for planets in the HZs of smaller host stellar targets, such as K and M dwarfs.

Works such as \citet{Cowan2012} have pointed out that a combination of thermal (IR) and reflected light (optical) observations would be necessary to fully characterize properties of exoplanets. As an exploration, we calculated the expected scale of phase variations in the optical and near-infrared by adopting the wide filter profiles of NIRCam on JWST \citep{nircam_doc}. The predicted scale is at best of order $10^{-10}$. Therefore, while reflected light variations hold interesting parallel physical observables, for the purposes and scope of this work our best case remains in the mid-infrared.

We also briefly explored the broad effect of ocean depth on our results, by re-running the $e=0.3$ cases with ocean mixed layer depths of 10 meters. The ice cover and cloud density increase slightly relative to the 50-meter case at both extremes of the orbit, for both rotation periods. The fast-rotation curves do exhibit stronger amplitudes of phase variations compared with their deeper ocean counterparts, but this variation is still small compared with the average difference between rotation cases at fixed depth. From this we conclude that while the ocean depth has some effect on the observables, both the quality and quantity of differences are not significant enough to affect our conclusions.
 
While we have attempted to explore a constrained problem with as few added assumptions as possible, we acknowledge that we have not considered other dynamical and atmospheric effects that might  cause large-amplitude variations in observables. In particular, we do not take into account the effects of tidal heating/dissipation, which could provide an additional forcing term to our systems. Additionally, the construction of transit and eclipse depths implies a narrow range of observed orbital inclinations, and we have assumed perfectly edge-on orbits in the construction of our predicted observables. This is a reflection of the detection bias inherent to transiting exoplanets; however, phase variations should persist even for non-transiting planets. We have also restricted ourselves to studying systems with solar-type host stars and mean instellations, even though studies such as \citet{yan13} predict that at instellations higher than Earth's, phase variations can effectively invert from the predictions for Earth-like instellation. Finally, we have assumed zero planetary obliquity; the interplay between the effects of the rotation and orbit in setting the periodicities in heating in particular would be greatly sensitive to the orientation of the spin axis. In future work we will examine how these results change for a variety of conditions for temperate terrestrial exoplanets.

\appendix

\section{Theoretical Estimates of Colors for Perfect Blackbodies}\label{sec:appendix:colors}
We constructed colors by taking the ratio of the planet-star contrast in one wavelength band to the planet-star contrast in a second wavelength band; we chose one band in a spectral region that is highly sensitive to water vapor absorption and the other in the water vapor window.  Here we illustrate the utility of these colors by examining the idealized case where the emission in each band comes from a perfect black body.  

\begin{figure}[htb!]
\begin{center}
   \begin{tabular}{cc}
     \includegraphics[width=7cm]{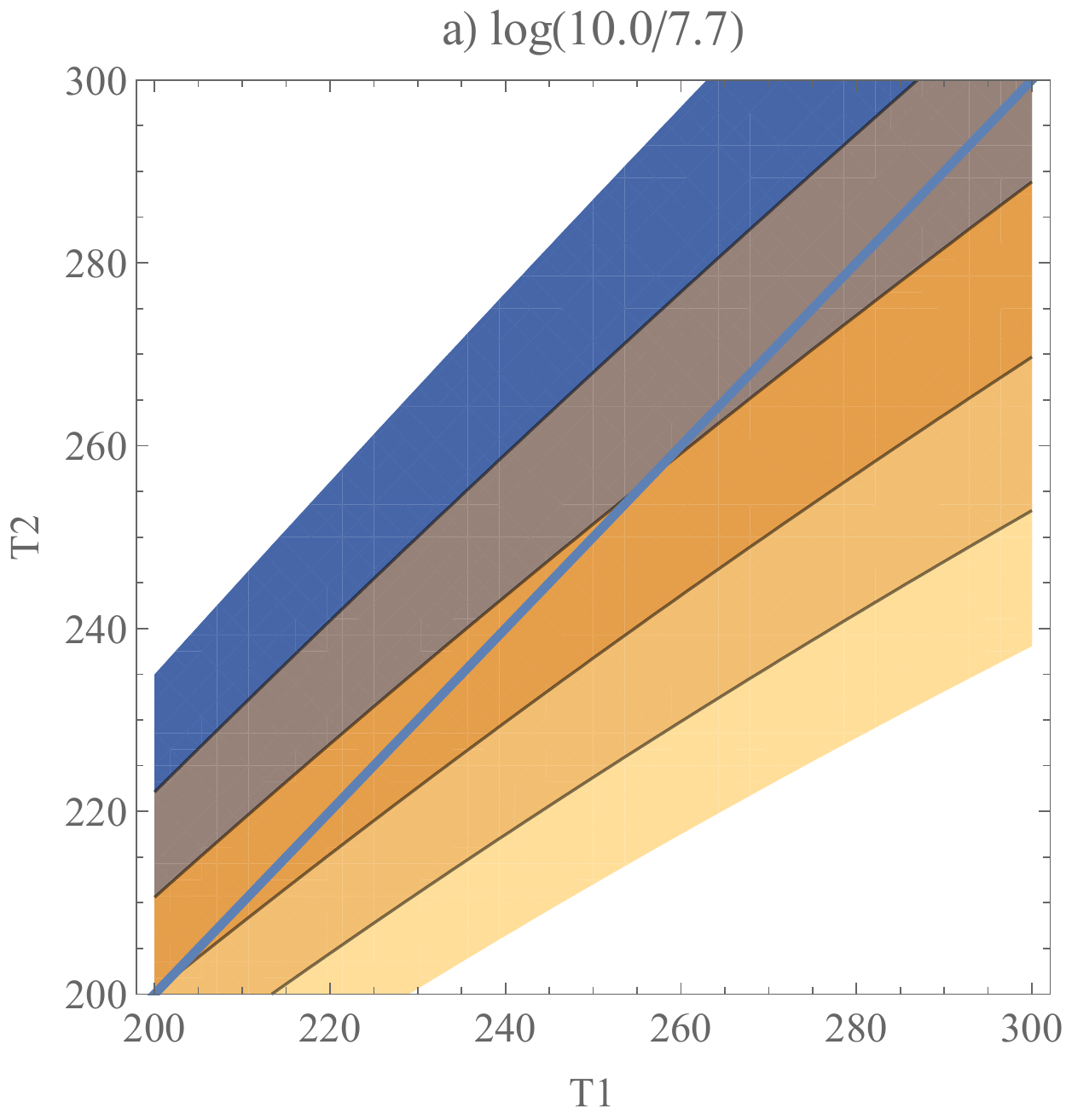}
     \includegraphics[width=1cm]{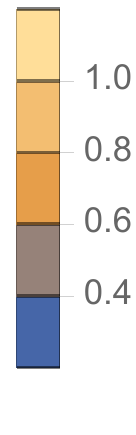} \\
     \includegraphics[width=7cm]{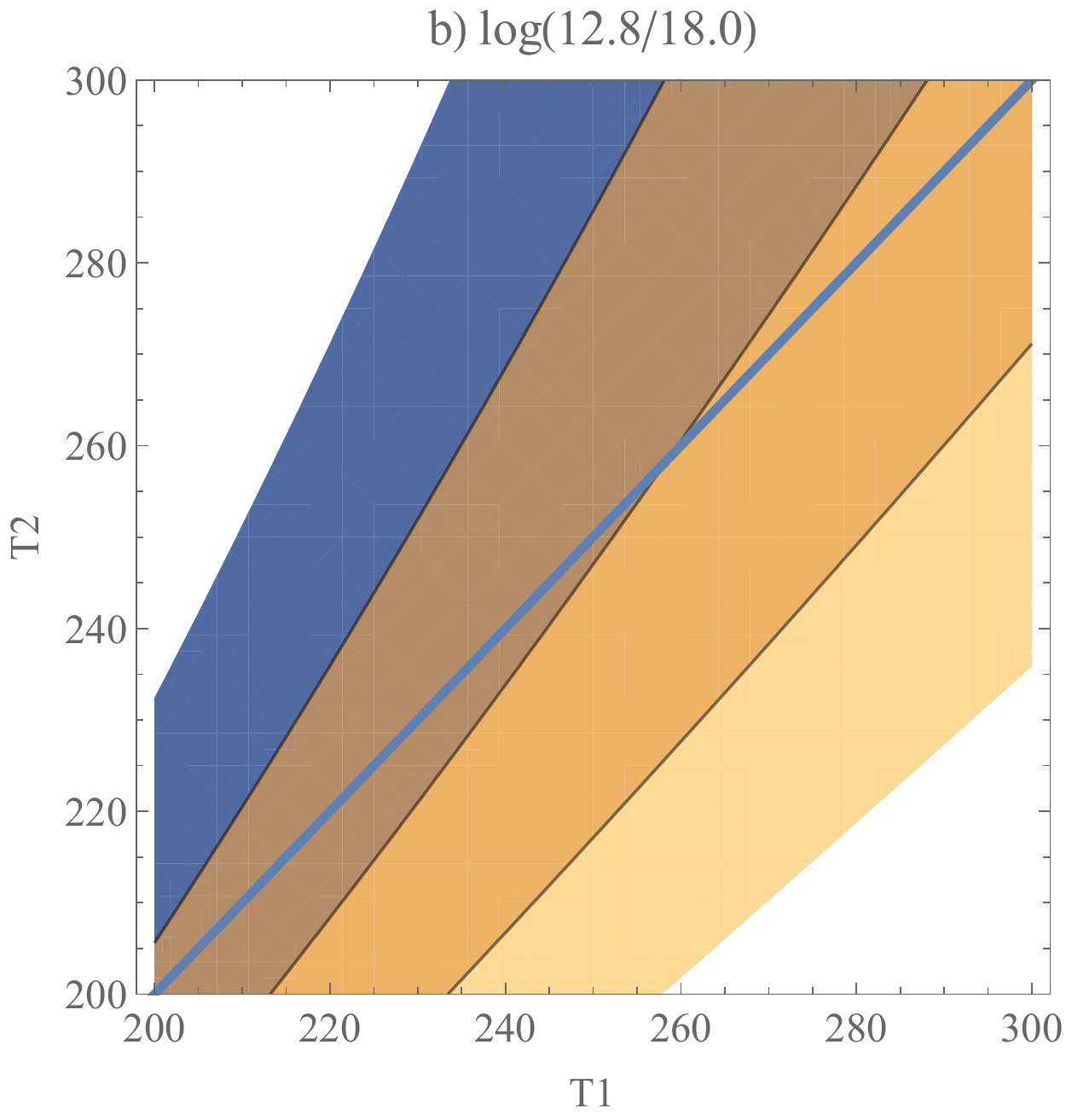} 
     \includegraphics[width=1cm]{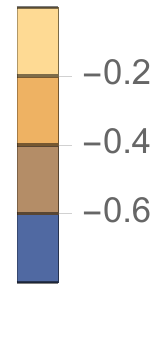} \\
   \end{tabular}
\caption{Theoretical ``colors'' for two perfect black bodies at temperatures $T_1$ and $T_2$, in Kelvin.  As described in the text, these colors are constructed using the Planck function and assuming a black body star at 6000 K, and use wavelengths of a) 10.0 and 7.7 $\mu$m, and b) 12.8 and 18.0 $\mu$m. The shading represents the values of the theoretical color given in the title of each panel, and the blue line shows where $T_1=T_2$.}
\label{fig:BlackbodyColors}
\end{center}
\end{figure}

In this idealized case, the planet-star contrast $P(\lambda,T)$ at wavelength $\lambda$ and planetary temperature $T$ is given in terms of the Planck function $B_{\lambda} (T)$,
\begin{equation}
P(\lambda,T) \propto \frac{B_\lambda(T)}{B_\lambda(T_\star)}
\end{equation}
where $T_\star$ is the emission temperature of the star, which we also approximate as a perfect black body.  The quantity $P(\lambda,T)$ is a thermal idealization of an individual light curve (e.g.\ Figure~\ref{fig:MIRI_lightcurves_omega90}). A color is then the ratio of the planet-star contrast at wavelength $\lambda_1$ to that at wavelength $\lambda_2$,
\begin{equation}
C(\lambda_1,\lambda_2) = \frac{P(\lambda_1,T_1)}{P(\lambda_2,T_2)} = \frac{B_{\lambda_1}(T_1) B_{\lambda_2}(T_\star)}{B_{\lambda_2}(T_2) B_{\lambda_1}(T_\star)}.
\end{equation}
We assume that the emission at wavelength $\lambda_i$ comes entirely from a layer of the atmosphere with temperature $T_i$, and we plot $C$ as a function of $T_1$, $T_2$ for two particular combinations of wavelengths (Figure~\ref{fig:BlackbodyColors}). For $(\lambda_1, \lambda_2)$ = (10.0 $\mu$m, 7.7 $\mu$m), lines of constant $C$ are slightly less steep than the one-to-one line. For a uniform warming of $T_1$ and $T_2$, an increase of 100 K is thus required to produce a decrease in $C$ of about 0.2 to 0.3 dex, while the same change in $C$ can be achieved by a differential warming in those temperatures of only 10--15 K. Similar behavior is exhibited for $(\lambda_1, \lambda_2)$ = (12.8 $\mu$m, 18.0 $\mu$m), except because of the reverse ordering of wavelengths, a uniform warming of 100 K produces an increase of about 0.2 dex, while that same increase can be achieved by a differential warming of 10--15 K.  Thus, $C$ is relatively insensitive to a uniform planetary warming or cooling, and is about an order of magnitude more sensitive to differential changes in the emission level of the two wavelengths chosen for $C$. For example, if one wavelength for $C$ is located in the water vapor window and the other in a spectral band with strong water vapor absorption, we would expect to see a large change in $C$ as we move from viewing a dry side of the planet (where $T_1$ and $T_2$ are nearly the same) to viewing a humid side. The emission changes due to the water-specific features are able to dominate the color variations over the orbit-induced thermal color variations.

\acknowledgements
This material is based upon work supported by the National Aeronautics and Space Administration through the NASA Astrobiology Institute under Cooperative Agreement Notice NNH13ZDA017C issued through the Science Mission Directorate. We acknowledge support from the NASA Astrobiology Institute through a cooperative agreement between NASA Ames Research Center and Yale University.

We would like to thank Dr. Aomawa Shields, whose feedback with regard to model convergence criteria and dependence of results on ocean mixed layer depth helped the scientific rigor of this work.

\software{Astropy \citep{ast13}, Colorcet \citep{kov15}, Jupyter \citep{klu16}, Matplotlib \citep{hun07}, Numpy \citep{van11}, Paletton \citep{pal18}, Scipy \citep{jon01}}

\clearpage
\bibliographystyle{aasjournal}
\bibliography{apj-jour,main}

\begin{thebibliography}{}
\expandafter\ifx\csname natexlab\endcsname\relax\def\natexlab#1{#1}\fi
\providecommand{\url}[1]{\href{#1}{#1}}
\providecommand{\dodoi}[1]{doi:~\href{http://doi.org/#1}{\nolinkurl{#1}}}
\providecommand{\doeprint}[1]{\href{http://ascl.net/#1}{\nolinkurl{http://ascl.net/#1}}}
\providecommand{\doarXiv}[1]{\href{https://arxiv.org/abs/#1}{\nolinkurl{https://arxiv.org/abs/#1}}}

\bibitem[{Adams \& Kane(2016)}]{ada16}
Adams, A.~D., \& Kane, S.~R. 2016, Astron. J., 152, 4,
  \dodoi{10.3847/0004-6256/152/1/4}

\bibitem[{Adams \& Laughlin(2018)}]{ada18b}
Adams, A.~D., \& Laughlin, G. 2018, Astron. J., 156, 28,
  \dodoi{10.3847/1538-3881/aac437}

\bibitem[{{Astropy Collaboration} {et~al.}(2013){Astropy Collaboration},
  Robitaille, Tollerud, Greenfield, Droettboom, Bray, Aldcroft, Davis,
  Ginsburg, Price-Whelan, Kerzendorf, Conley, Crighton, Barbary, Muna,
  Ferguson, Grollier, Parikh, Nair, Unther, Deil, Woillez, Conseil, Kramer,
  Turner, Singer, Fox, Weaver, Zabalza, Edwards, {Azalee Bostroem}, Burke,
  Casey, Crawford, Dencheva, Ely, Jenness, Labrie, Lim, Pierfederici, Pontzen,
  Ptak, Refsdal, Servillat, \& Streicher}]{ast13}
{Astropy Collaboration}, Robitaille, T., Tollerud, E., {et~al.} 2013, \aap,
  558, A33, \dodoi{10.1051/0004-6361/201322068}

\bibitem[{Bardeen {et~al.}(2017)Bardeen, Garcia, Toon, \& Conley}]{Bardeen2017}
Bardeen, C.~G., Garcia, R.~R., Toon, O.~B., \& Conley, A.~J. 2017, Proc. Natl.
  Acad. Sci., 114, E7415, \dodoi{10.1073/pnas.1708980114}

\bibitem[{Barnes {et~al.}(1996)Barnes, Haberle, Pollack, Lee, \&
  Schaeffer}]{bar96a}
Barnes, J., Haberle, R., Pollack, J., Lee, H., \& Schaeffer, J. 1996, \jgr,
  101, 12753, \dodoi{10.1029/96JE00179}

\bibitem[{Barnes {et~al.}(1993)Barnes, Pollack, Haberle, Leovy, Zurek, Lee, \&
  Schaeffer}]{bar93}
Barnes, J., Pollack, J., Haberle, R., {et~al.} 1993, \jgr, 98, 3125,
  \dodoi{10.1029/92JE02935}

\bibitem[{Barnes \& Haberle(1996)}]{bar96b}
Barnes, J.~R., \& Haberle, R.~M. 1996, J. Atmos. Sci., 53, 3143,
  \dodoi{10.1175/1520-0469(1996)053<3143:TMZMCA>2.0.CO;2}

\bibitem[{Barnes {et~al.}(2008)Barnes, Raymond, Jackson, \& Greenberg}]{bar08}
Barnes, R., Raymond, S., Jackson, B., \& Greenberg, R. 2008, Astrobiology, 8,
  557, \dodoi{10.1089/ast.2007.0204}

\bibitem[{Battersby {et~al.}(2018)Battersby, Armus, Bergin, Kataria, Meixner,
  Pope, Stevenson, Cooray, Leisawitz, Scott, Bauer, Bradford, Ennico, Fortney,
  Kaltenegger, Melnick, Milam, Narayanan, Padgett, Pontoppidan, Roellig,
  Sandstrom, Su, Vieira, Wright, Zmuidzinas, Staguhn, Sheth, Benford, Mamajek,
  Neff, Carey, Burgarella, {De Beck}, Gerin, Helmich, Moseley, Sakon, \&
  Wiedner}]{Battersby2018}
Battersby, C., Armus, L., Bergin, E., {et~al.} 2018, Nat. Astron., 2, 596,
  \dodoi{10.1038/s41550-018-0540-y}

\bibitem[{Bengtsson {et~al.}(2013)Bengtsson, Bonnet, Grinspoon, Koumoutsaris,
  Lebonnois, \& Titov}]{Venus2013}
Bengtsson, L., Bonnet, R.-M., Grinspoon, D., {et~al.}, eds. 2013, {Towards
  Understanding the Climate of Venus} (New York, NY: Springer New York),
  \dodoi{10.1007/978-1-4614-5064-1}.
\newblock \url{http://link.springer.com/10.1007/978-1-4614-5064-1}

\bibitem[{Bitz {et~al.}(2012)Bitz, Shell, Gent, Bailey, Danabasoglu, Armour,
  Holland, \& Kiehl}]{Bitz2012}
Bitz, C.~M., Shell, K.~M., Gent, P.~R., {et~al.} 2012, J. Clim., 25, 3053,
  \dodoi{10.1175/JCLI-D-11-00290.1}

\bibitem[{Bolmont {et~al.}(2016)Bolmont, Libert, Leconte, \& Selsis}]{bol16}
Bolmont, E., Libert, A.-S. A.-S., Leconte, J., \& Selsis, F. 2016, \aap, 591,
  A106, \dodoi{10.1051/0004-6361/201628073}

\bibitem[{Brent(1973)}]{Brent1973}
Brent, R.~P. 1973, {Algorithms for Minimization without Derivatives} (Englewood
  Cliffs, NJ: Prentice-Hall), 195

\bibitem[{Chiang \& Bitz(2005)}]{Chiang2005}
Chiang, J. C.~H., \& Bitz, C.~M. 2005, Clim. Dyn., 25, 477,
  \dodoi{10.1007/s00382-005-0040-5}

\bibitem[{Cowan \& Agol(2011)}]{cow11a}
Cowan, N., \& Agol, E. 2011, \apj, 726, 82, \dodoi{10.1088/0004-637X/726/2/82}

\bibitem[{Cowan {et~al.}(2012)Cowan, Voigt, \& Abbot}]{Cowan2012}
Cowan, N.~B., Voigt, A., \& Abbot, D.~S. 2012, Astrophys. J., 757,
  \dodoi{10.1088/0004-637X/757/1/80}

\bibitem[{Cullum {et~al.}(2014)Cullum, Stevens, \& Joshi}]{Cullum2014}
Cullum, J., Stevens, D., \& Joshi, M. 2014, Astrobiology, 14, 645,
  \dodoi{10.1089/ast.2014.1171}

\bibitem[{de~Wit {et~al.}(2016)de~Wit, Lewis, Langton, Laughlin, Deming,
  Batygin, \& Fortney}]{dew16}
de~Wit, J., Lewis, N.~K., Langton, J., {et~al.} 2016, \apjl, 820, L33,
  \dodoi{10.3847/2041-8205/820/2/L33}

\bibitem[{Deitrick {et~al.}(2018)Deitrick, Barnes, Bitz, Fleming, Charnay,
  Wilhelm, Armstrong, Quinn, \& Meadows}]{Deitrick2018}
Deitrick, R., Barnes, R., Bitz, C., {et~al.} 2018, Astron. J., 155, 266,
  \dodoi{10.3847/1538-3881/aac214}

\bibitem[{Forget {et~al.}(2013)Forget, Wordsworth, Millour, Madeleine, Kerber,
  Leconte, Marcq, \& Haberle}]{for13}
Forget, F., Wordsworth, R., Millour, E., {et~al.} 2013, \icarus, 222, 81,
  \dodoi{10.1016/j.icarus.2012.10.019}

\bibitem[{Forget {et~al.}(1999)Forget, Hourdin, Fournier, Hourdin, Talagrand,
  Collins, Lewis, Read, \& Huot}]{for99}
Forget, F., Hourdin, F., Fournier, R., {et~al.} 1999, \jgr, 104, 24155,
  \dodoi{10.1029/1999JE001025}

\bibitem[{Fortney {et~al.}(2018)Fortney, Kataria, Stevenson, Zellem, Nielsen,
  Cuartas-Restrepo, Gaidos, Bergin, Meixner, Kane, David, Fraine, Kaltenegger,
  Tanner, Lopez-Morales, Greene, Danchi, Stassun, Kopparapu, Wolf, Meshkat,
  Hinkel, Pontoppidan, Dong, Bruno, Gelino, Airapetian, Agol, Deming,
  Haqq-Misra, Parenteau, Lisse, Tucker, Saxena, Wordsworth, Blake, Curry,
  Berta-Thompson, Fridlund, Su, Gao, Adibekyan, Heavens, Minniti, Rugheimer,
  Rackham, Mandt, de~Val-Borro, \& Robinson}]{Fortney2018}
Fortney, J., Kataria, T., Stevenson, K., {et~al.} 2018.
\newblock \doarXiv{1803.07730}

\bibitem[{Georgakarakos {et~al.}(2016)Georgakarakos, Dobbs-Dixon, \&
  Way}]{Georgakarakos2016}
Georgakarakos, N., Dobbs-Dixon, I., \& Way, M.~J. 2016, Mon. Not. R. Astron.
  Soc., 461, 1512, \dodoi{10.1093/mnras/stw1378}

\bibitem[{Glasse {et~al.}(2015)Glasse, Rieke, Bauwens, Garc\'{i}a-Mar\'{i}n,
  Ressler, Rost, Tikkanen, Vandenbussche, \& Wright}]{gla15}
Glasse, A., Rieke, G., Bauwens, E., {et~al.} 2015, \pasp, 127, 686,
  \dodoi{10.1086/682259}

\bibitem[{Goldreich \& Soter(1966)}]{gol66}
Goldreich, P., \& Soter, S. 1966, Icarus, 5, 375,
  \dodoi{http://dx.doi.org/10.1016/0019-1035(66)90051-0}

\bibitem[{Haberle {et~al.}(1993)Haberle, Pollack, Barnes, Zurek, Leovy, Murphy,
  Lee, \& Schaeffer}]{hab93}
Haberle, R., Pollack, J., Barnes, J., {et~al.} 1993, \jgr, 98, 3093,
  \dodoi{10.1029/92JE02946}

\bibitem[{Hack(1994)}]{Hack1994}
Hack, J.~J. 1994, J. Geophys. Res., 99, 5551, \dodoi{10.1029/93JD03478}

\bibitem[{{Hunke E.C.}(2008)}]{hun08}
{Hunke E.C.}, L. W.~H. 2008, {CICE: The Los Alamos Sea Ice Model, Documentation
  and Software User's Manual, Version 4.0}, Tech. Rep. LA-CC-06-012, Los Alamos
  National Laboratory, Los Alamos, New Mexico

\bibitem[{Hunter(2007)}]{hun07}
Hunter, J.~D. 2007, Comput. Sci. Eng., 9, 90, \dodoi{10.1109/MCSE.2007.55}

\bibitem[{Hut(1981)}]{hut81}
Hut, P. 1981, \aap, 99, 126, \dodoi{1981A&A....99..126H}

\bibitem[{Jones {et~al.}(2001)Jones, Oliphant, Peterson, \& Others}]{jon01}
Jones, E., Oliphant, T., Peterson, P., \& Others. 2001, {{SciPy}: Open source
  scientific tools for {Python}}.
\newblock \url{http://www.scipy.org/}

\bibitem[{Joshi {et~al.}(1997)Joshi, Haberle, \& Reynolds}]{jos97}
Joshi, M., Haberle, R., \& Reynolds, R. 1997, \icarus, 129, 450,
  \dodoi{10.1006/icar.1997.5793}

\bibitem[{Kasting {et~al.}(1993)Kasting, Whitmire, \& Reynolds}]{kas93}
Kasting, J., Whitmire, D., \& Reynolds, R. 1993, \icarus, 101, 108,
  \dodoi{10.1006/icar.1993.1010}

\bibitem[{Kataria(2018)}]{Kataria2018}
Kataria, T. 2018, in COSPAR Meeting, Vol.~42, 42nd COSPAR Sci. Assem.,
  E4.1--18--18

\bibitem[{Kataria {et~al.}(2013)Kataria, Showman, Lewis, Fortney, Marley, \&
  Freedman}]{Kataria2013}
Kataria, T., Showman, A., Lewis, N., {et~al.} 2013, \apj, 767, 76,
  \dodoi{10.1088/0004-637X/767/1/76}

\bibitem[{Kluyver {et~al.}(2016)Kluyver, Ragan-Kelley, Perez, Granger,
  Bussonnier, Frederic, Kelley, Hamrick, Grout, Corlay, Ivanov, Avila, Abdalla,
  Willing, \& Team}]{klu16}
Kluyver, T., Ragan-Kelley, B., Perez, F., {et~al.} 2016, IOS Press, 87

\bibitem[{Kopparapu {et~al.}(2014)Kopparapu, Ramirez, SchottelKotte, Kasting,
  Domagal-Goldman, \& Eymet}]{kop14}
Kopparapu, R., Ramirez, R., SchottelKotte, J., {et~al.} 2014, \apjl, 787, L29,
  \dodoi{10.1088/2041-8205/787/2/L29}

\bibitem[{Kopparapu {et~al.}(2017)Kopparapu, Wolf, Arney, Batalha, Haqq-Misra,
  Grimm, \& Heng}]{Kopparapu2017}
Kopparapu, R., Wolf, E.~T., Arney, G., {et~al.} 2017, Astrophys. J., 845, 5,
  \dodoi{10.3847/1538-4357/aa7cf9}

\bibitem[{Kopparapu {et~al.}(2013)Kopparapu, Ramirez, Kasting, Eymet, Robinson,
  Mahadevan, Terrien, Domagal-Goldman, Meadows, \& Deshpande}]{kop13}
Kopparapu, R., Ramirez, R., Kasting, J., {et~al.} 2013, \apj, 765, 131,
  \dodoi{10.1088/0004-637X/765/2/131}

\bibitem[{Kovesi(2015)}]{kov15}
Kovesi, P. 2015, ArXiv e-prints.
\newblock \doarXiv{1509.03700}

\bibitem[{Kozai(1962)}]{koz62}
Kozai, Y. 1962, \aj, 67, 591, \dodoi{10.1086/108790}

\bibitem[{Kreidberg(2018)}]{Kreidberg2018}
Kreidberg, L. 2018, in Handb. Exopl. (Cham: Springer International Publishing),
  1--23.
\newblock \url{http://link.springer.com/10.1007/978-3-319-30648-3_100-1}

\bibitem[{Langton \& Laughlin(2008)}]{Langton2008}
Langton, J., \& Laughlin, G. 2008, Astrophys. J., 674, 1106,
  \dodoi{10.1086/523957}

\bibitem[{Lebonnois {et~al.}(2010)Lebonnois, Hourdin, Eymet, Crespin, Fournier,
  \& Forget}]{leb10}
Lebonnois, S., Hourdin, F., Eymet, V., {et~al.} 2010, J. Geophys. Res., 115,
  E06006, \dodoi{10.1029/2009JE003458}

\bibitem[{Leconte {et~al.}(2013)Leconte, Forget, Charnay, Wordsworth, Selsis,
  Millour, \& Spiga}]{Leconte2013}
Leconte, J., Forget, F., Charnay, B., {et~al.} 2013, Astron. Astrophys., 554,
  A69, \dodoi{10.1051/0004-6361/201321042}

\bibitem[{Lewis {et~al.}(2017)Lewis, Parmentier, Kataria, de~Wit, Showman,
  Fortney, \& Marley}]{lew17}
Lewis, N., Parmentier, V., Kataria, T., {et~al.} 2017, ArXiv e-prints.
\newblock \doarXiv{1706.00466}

\bibitem[{Lidov(1962)}]{lid62}
Lidov, M. 1962, \planss, 9, 719, \dodoi{10.1016/0032-0633(62)90129-0}

\bibitem[{Lin \& Rood(1997)}]{lin97}
Lin, S., \& Rood, R. 1997, Q. J. R. Meteorol. Soc., 123, 2477,
  \dodoi{10.1002/qj.49712354416}

\bibitem[{Lin \& Rood(1996)}]{lin96}
Lin, S.-J., \& Rood, R. 1996, Mon. Weather Rev., 124, 2046,
  \dodoi{10.1175/1520-0493(1996)124<2046:MFFSLT>2.0.CO;2}

\bibitem[{Linsenmeier {et~al.}(2015)Linsenmeier, Pascale, \& Lucarini}]{lin15}
Linsenmeier, M., Pascale, S., \& Lucarini, V. 2015, \planss, 105, 43,
  \dodoi{10.1016/j.pss.2014.11.003}

\bibitem[{Makarov \& Efroimsky(2013)}]{mak13}
Makarov, V.~V., \& Efroimsky, M. 2013, Astrophys. J., 764, 27,
  \dodoi{10.1088/0004-637X/764/1/27}

\bibitem[{Merlis \& Schneider(2010)}]{mer10}
Merlis, T., \& Schneider, T. 2010, J. Adv. Model. Earth Syst., 2, 13,
  \dodoi{10.3894/JAMES.2010.2.13}

\bibitem[{Murray \& Dermott(1999)}]{1999ssd..book.....M}
Murray, C., \& Dermott, S. 1999, {Solar system dynamics}

\bibitem[{Neale {et~al.}(2010)Neale, Chen, Gettelman, Lauritzen, Park, \&
  Williamson}]{nea10}
Neale, R.~B., Chen, C.-C., Gettelman, A., {et~al.} 2010, {Description of the
  NCAR community atmosphere model (CAM 5.0)}, Tech. rep.

\bibitem[{O'Dell {et~al.}(2008)O'Dell, Wentz, \& Bennartz}]{ODell2008}
O'Dell, C.~W., Wentz, F.~J., \& Bennartz, R. 2008, J. Clim., 21, 1721,
  \dodoi{10.1175/2007JCLI1958.1}

\bibitem[{O'Gorman \& Schneider(2008)}]{ogo08}
O'Gorman, P., \& Schneider, T. 2008, J. Clim., 21, 3815,
  \dodoi{10.1175/2007JCLI2065.1}

\bibitem[{Parmentier \& Crossfield(2018)}]{par17}
Parmentier, V., \& Crossfield, I. J.~M. 2018, in Handb. Exopl. (Cham: Springer
  International Publishing), 1--22.
\newblock \url{http://arxiv.org/abs/1711.07696
  http://dx.doi.org/10.1007/978-3-319-30648-3_116-1
  http://link.springer.com/10.1007/978-3-319-30648-3_116-1}

\bibitem[{Paynter \& Ramaswamy(2011)}]{Paynter2011}
Paynter, D.~J., \& Ramaswamy, V. 2011, J. Geophys. Res., 116, D20302,
  \dodoi{10.1029/2010JD015505}

\bibitem[{Pierrehumbert \& Hammond(2019)}]{Pierrehumbert2019}
Pierrehumbert, R.~T., \& Hammond, M. 2019, Annu. Rev. Fluid Mech., 51, 275,
  \dodoi{10.1146/annurev-fluid-010518-040516}

\bibitem[{Pincus {et~al.}(2003)Pincus, Barker, \& Morcrette}]{Pincus2003}
Pincus, R., Barker, H.~W., \& Morcrette, J.-J. 2003, J. Geophys. Res. Atmos.,
  108, n/a, \dodoi{10.1029/2002JD003322}

\bibitem[{Ramirez \& Kaltenegger(2016)}]{Ramirez2016}
Ramirez, R., \& Kaltenegger, L. 2016, Astrophys. J., 823, 1,
  \dodoi{10.3847/0004-637X/823/1/6}

\bibitem[{Ramirez(2018)}]{Ramirez2018}
Ramirez, R.~M. 2018, 1, \dodoi{10.3390/geosciences8080280}

\bibitem[{Rasch \& Kristj{\'{a}}nsson(1998)}]{Rasch1998}
Rasch, P.~J., \& Kristj{\'{a}}nsson, J.~E. 1998, J. Clim., 11, 1587,
  \dodoi{10.1175/1520-0442(1998)011<1587:ACOTCM>2.0.CO;2}

\bibitem[{Raymond \& Blyth(1986)}]{Raymond1986}
Raymond, D.~J., \& Blyth, A.~M. 1986, J. Atmos. Sci., 43, 2708,
  \dodoi{10.1175/1520-0469(1986)043<2708:ASMMFN>2.0.CO;2}

\bibitem[{Raymond \& Blyth(1992)}]{Raymond1992}
---. 1992, J. Atmos. Sci., 49, 1968,
  \dodoi{10.1175/1520-0469(1992)049<1968:EOTSMM>2.0.CO;2}

\bibitem[{Rieke {et~al.}(2015)Rieke, Wright, B{\"{o}}ker, Bouwman, Colina,
  Glasse, Gordon, Greene, G{\"{u}}del, Henning, Justtanont, Lagage, Meixner,
  N{\o}rgaard-Nielsen, Ray, Ressler, van Dishoeck, \& Waelkens}]{rie15}
Rieke, G., Wright, G., B{\"{o}}ker, T., {et~al.} 2015, \pasp, 127, 584,
  \dodoi{10.1086/682252}

\bibitem[{Rossow(1983)}]{ros83}
Rossow, W. 1983, J. Atmos. Sci., 40, 273,
  \dodoi{10.1175/1520-0469(1983)040<0273:AGCMOA>2.0.CO;2}

\bibitem[{Sakon {et~al.}(2018)Sakon, Roellig, Ennico-Smith, Ikeda, Matsuo,
  Yamamuro, Fujishiro, Enya, Kotani, Nishikawa, Sarugaku, Takahashi, Wada,
  Burgarella, Murakami, \& Guyon}]{Sakon2018}
Sakon, I., Roellig, T.~L., Ennico-Smith, K., {et~al.} 2018, in Sp. Telesc.
  Instrum. 2018 Opt. Infrared, Millim. Wave, ed. H.~A. MacEwen, M.~Lystrup,
  G.~G. Fazio, N.~Batalha, E.~C. Tong, \& N.~Siegler (SPIE), 42.
\newblock
  \url{https://www.spiedigitallibrary.org/conference-proceedings-of-spie/10698/2314177/The-mid-infrared-imager-spectrometer-coronagraph-MISC-for-the-Origins/10.1117/12.2314177.full}

\bibitem[{Schneider(2006)}]{sch06}
Schneider, T. 2006, Annu. Rev. Earth Planet. Sci., 34, 655,
  \dodoi{10.1146/annurev.earth.34.031405.125144}

\bibitem[{Selsis {et~al.}(2013)Selsis, Maurin, Hersant, Leconte, Bolmont,
  Raymond, \& Delbo}]{Selsis2013}
Selsis, F., Maurin, A.~S., Hersant, F., {et~al.} 2013, Orig. Life Evol.
  Biosph., 41, 539, \dodoi{10.1051/0004-6361/201321661}

\bibitem[{Shields {et~al.}(2014)Shields, Bitz, Meadows, Joshi, \&
  Robinson}]{shi14}
Shields, A., Bitz, C., Meadows, V., Joshi, M., \& Robinson, T. 2014, \apjl,
  785, L9, \dodoi{10.1088/2041-8205/785/1/L9}

\bibitem[{Shields {et~al.}(2013)Shields, Meadows, Bitz, Pierrehumbert, Joshi,
  \& Robinson}]{shi13}
Shields, A., Meadows, V., Bitz, C., {et~al.} 2013, Astrobiology, 13, 715,
  \dodoi{10.1089/ast.2012.0961}

\bibitem[{Shields {et~al.}(2016)Shields, Barnes, Agol, Charnay, Bitz, \&
  Meadows}]{shi16}
Shields, A. L.~A., Barnes, R., Agol, E., {et~al.} 2016, Astrobiology, 16, 443,
  \dodoi{10.1089/ast.2015.1353}

\bibitem[{Showman {et~al.}(2010)Showman, Cho, \& Menou}]{Showman2010}
Showman, A., Cho, J.-K., \& Menou, K. 2010, {Atmospheric Circulation of
  Exoplanets}, ed. S.~Seager, 471--516

\bibitem[{Showman \& Guillot(2002{\natexlab{a}})}]{sho02}
Showman, A., \& Guillot, T. 2002{\natexlab{a}}, \aap, 385, 166,
  \dodoi{10.1051/0004-6361:20020101}

\bibitem[{Showman \& Guillot(2002{\natexlab{b}})}]{Showman2002}
Showman, A.~P., \& Guillot, T. 2002{\natexlab{b}}, Astron. Astrophys., 385,
  166, \dodoi{10.1051/0004-6361:20020101}

\bibitem[{{Space Telescope Science Institute}(2017)}]{nircam_doc}
{Space Telescope Science Institute}. 2017, {NIRCam Imaging},  Baltimore, MD:
  JWST User Documentation [Published 2017 July 13].
\newblock \url{https://jwst-docs.stsci.edu/display/JTI/NIRCam+Imaging}

\bibitem[{{Space Telescope Science Institute}(2018)}]{miri_doc}
---. 2018, {MIRI Imaging},  Baltimore, MD: JWST User Documentation [Published
  2018 February 9].
\newblock \url{https://jwst-docs.stsci.edu/display/JTI/MIRI+Overview}

\bibitem[{Spiegel {et~al.}(2010)Spiegel, Raymond, Dressing, Scharf, \&
  Mitchell}]{spi10}
Spiegel, D., Raymond, S., Dressing, C., Scharf, C., \& Mitchell, J. 2010, \apj,
  721, 1308, \dodoi{10.1088/0004-637X/721/2/1308}

\bibitem[{Stan\'{i}{\v{c}}ek(2018)}]{pal18}
Stan\'{i}{\v{c}}ek, P. 2018, {Paletton, the color scheme designer}.
\newblock \url{http://paletton.com/}

\bibitem[{Toon {et~al.}(1989)Toon, McKay, Ackerman, \& Santhanam}]{Toon1989}
Toon, O.~B., McKay, C.~P., Ackerman, T.~P., \& Santhanam, K. 1989, J. Geophys.
  Res., 94, 16287, \dodoi{10.1029/JD094iD13p16287}

\bibitem[{van~der Walt {et~al.}(2011)van~der Walt, Colbert, \&
  Varoquaux}]{van11}
van~der Walt, S., Colbert, S.~C., \& Varoquaux, G. 2011, Comput. Sci. Eng., 13,
  22, \dodoi{10.1109/MCSE.2011.37}

\bibitem[{Way {et~al.}(2018)Way, {Del Genio}, Aleinov, Clune, Kelley, \&
  Kiang}]{Way2018}
Way, M.~J., {Del Genio}, A.~D., Aleinov, I., {et~al.} 2018, Astrophys. J.
  Suppl. Ser., 239, 24, \dodoi{10.3847/1538-4365/aae9e1}

\bibitem[{Way \& Georgakarakos(2016)}]{Way2016}
Way, M.~J., \& Georgakarakos, N. 2016, Astrophys. J. Lett., 835, 1,
  \dodoi{10.3847/2041-8213/835/1/L1}

\bibitem[{Werner {et~al.}(2004)Werner, Roellig, Low, Rieke, Rieke, Hoffmann,
  Young, Houck, Brandl, Fazio, Hora, Gehrz, Helou, Soifer, Stauffer, Keene,
  Eisenhardt, Gallagher, Gautier, Irace, Lawrence, Simmons, {Van Cleve}, Jura,
  Wright, \& Cruikshank}]{wer04}
Werner, M., Roellig, T., Low, F., {et~al.} 2004, \apjs, 154, 1,
  \dodoi{10.1086/422992}

\bibitem[{Williams \& Pollard(2002)}]{wil02}
Williams, D.~M., \& Pollard, D. 2002, Int. J. Astrobiol., 1, 61,
  \dodoi{10.1017/S1473550402001064}

\bibitem[{Wolf \& Toon(2013)}]{wol13}
Wolf, E., \& Toon, O. 2013, Astrobiology, 13, 656,
  \dodoi{10.1089/ast.2012.0936}

\bibitem[{Wolf \& Toon(2014)}]{wol14}
---. 2014, \grl, 41, 167, \dodoi{10.1002/2013GL058376}

\bibitem[{Wolf \& Toon(2015)}]{wol15}
---. 2015, J. Geophys. Res., 120, 5775, \dodoi{10.1002/2015JD023302}

\bibitem[{Wright {et~al.}(2015)Wright, Wright, Goodson, Rieke, Aitink-Kroes,
  Amiaux, Aricha-Yanguas, Azzollini, Banks, Barrado-Navascues,
  Belenguer-Davila, Bloemmart, Bouchet, Brandl, Colina, Detre, Diaz-Catala,
  Eccleston, Friedman, Garc\'{i}a-Mar\'{i}n, G{\"{u}}del, Glasse, Glauser,
  Greene, Groezinger, Grundy, Hastings, Henning, Hofferbert, Hunter, Jessen,
  Justtanont, Karnik, Khorrami, Krause, Labiano, Lagage, Langer, Lemke, Lim,
  Lorenzo-Alvarez, Mazy, McGowan, Meixner, Morris, Morrison, M{\"{u}}ller,
  Rgaard-Nielson, Olofsson, O'Sullivan, Pel, Penanen, Petach, Pye, Ray,
  Renotte, Renouf, Ressler, Samara-Ratna, Scheithauer, Schneider, Shaughnessy,
  Stevenson, Sukhatme, Swinyard, Sykes, Thatcher, Tikkanen, van Dishoeck,
  Waelkens, Walker, Wells, \& Zhender}]{wri15}
Wright, G., Wright, D., Goodson, G., {et~al.} 2015, \pasp, 127, 595,
  \dodoi{10.1086/682253}

\bibitem[{Yang {et~al.}(2013)Yang, Cowan, \& Abbot}]{yan13}
Yang, J., Cowan, N. B.~N., \& Abbot, D. D.~S. 2013, \apjl, 771, L45,
  \dodoi{10.1088/2041-8205/771/2/L45}

\bibitem[{Zalucha {et~al.}(2010)Zalucha, Plumb, \& Wilson}]{zal10}
Zalucha, A., Plumb, R., \& Wilson, R. 2010, J. Atmos. Sci., 67, 673,
  \dodoi{10.1175/2009JAS3130.1}

\bibitem[{Zhang \& McFarlane(1995)}]{Zhang1995}
Zhang, G.~J., \& McFarlane, N.~A. 1995, J. Geophys. Res. Atmos., 100, 1417,
  \dodoi{10.1029/94JD02519}

\end{thebibliography}

\end{document}